\documentclass[12pt]{article}
\usepackage{amsmath,amssymb}
\usepackage{graphicx}
\newcommand{\eqdef}{\stackrel{\text{def}}{=}}
\newcommand{\n}{\nonumber\\}
\newcommand{\bm}{\boldsymbol}
\newcommand{\cF}{c_{\text{\tiny$\mathcal{F}$}}}
\newcommand{\ignore}[1]{}
\numberwithin{equation}{section}
\newcommand{\Romannumeral}[1]{\uppercase\expandafter{\romannumeral#1}}
\newcommand{\I}{\text{\Romannumeral{1}}}
\newcommand{\II}{\text{\Romannumeral{2}}}

\allowdisplaybreaks[4]

\begin{document}

\baselineskip=20pt

\newfont{\elevenmib}{cmmib10 scaled\magstep1}
\newcommand{\preprint}{
   \begin{flushright}\normalsize \sf
     DPSU-13-1\\
   \end{flushright}}
\newcommand{\Title}[1]{{\baselineskip=26pt
   \begin{center} \Large \bf #1 \\ \ \\ \end{center}}}
\newcommand{\Author}{\begin{center}
   \large \bf Satoru Odake \end{center}}
\newcommand{\Address}{\begin{center}
     Department of Physics, Shinshu University,\\
     Matsumoto 390-8621, Japan
   \end{center}}
\newcommand{\Accepted}[1]{\begin{center}
   {\large \sf #1}\\ \vspace{1mm}{\small \sf Accepted for Publication}
   \end{center}}

\preprint
\thispagestyle{empty}

\Title{Recurrence Relations of\\ the Multi-Indexed Orthogonal Polynomials}

\Author

\Address
\vspace{1cm}

\begin{abstract}
Ordinary orthogonal polynomials are uniquely characterized by the three term
recurrence relations up to an overall multiplicative constant.
We show that the newly discovered $M$-indexed orthogonal polynomials satisfy
$3+2M$ term recurrence relations with non-trivial initial data of the lowest
$M+1$ members. These include the multi-indexed orthogonal polynomials of
Laguerre, Jacobi, Wilson and Askey-Wilson types.
The $M=0$ case is the corresponding classical orthogonal polynomials.
\end{abstract}

\section{Introduction}
\label{intro}

Exactly solvable quantum mechanical systems in one dimension had been
well studied \cite{infhul}--\cite{susyqm}. In recent years the subject
saw remarkable developments \cite{gomez}--\cite{os27}.
The eigenfunctions of the solvably deformed systems are described by
new types of orthogonal polynomials, the {\em exceptional} and
{\em multi-indexed orthogonal polynomials}.
The first breakthrough was the discovery of the $X_1$ Laguerre and Jacobi
exceptional orthogonal polynomials in the context of Strum-Liouville
theory by G\'{o}mez-Ullate, Kamran and Milson \cite{gomez}.
Quesne constructed shape-invariant quantum mechanical systems whose
eigenfunctions are described by the $X_1$ Laguerre and Jacobi polynomials
\cite{quesne}.
The second progress was the construction of the $X_{\ell}$ Laguerre and
Jacobi exceptional orthogonal polynomials for any positive integer $\ell$
by Sasaki and the present author \cite{os16}, which were based on the
deformations of the quantum mechanical systems preserving the
shape-invariance \cite{genden}.
The third development was the generalization of the exceptional orthogonal
polynomials, i.e., multi-indexed orthogonal polynomials of Laguerre and
Jacobi types, which were obtained based on the method of virtual states
deletion for quantum mechanical systems \cite{gomez3,os25}.
The exceptional orthogonal polynomial (in the narrow sense) is a
one-indexed orthogonal polynomial.
Parallel to the ordinary quantum mechanical systems, the discrete quantum
mechanical systems had been developed \cite{os15}--\cite{os24} and
the exceptional and multi-indexed orthogonal polynomials of Wilson,
Askey-Wilson, Racah and $q$-Racah types were constructed
\cite{os17}--\cite{os27}.

By the method of virtual states deletion \cite{os25}, which is based on
the Darboux-Crum transformation \cite{darb,crum}, infinitely many exactly
solvable quantum mechanical systems are systematically obtained from
the original exactly solvable systems.
The multi-indexed orthogonal polynomials of Laguerre, Jacobi, Wilson,
Askey-Wilson, Racah and $q$-Racah types describe their eigenfunctions.
They satisfy second order differential or difference equations and
form complete basis but their degrees start at a certain positive integer
$\ell$ instead of zero, namely the set of degrees is
$\{\ell,\ell+1,\ell+2,\ldots\}$ (a maximum degree exists for ($q$-)Racah cases).
Thus the constraints of Bochner's theorem \cite{bochner} are avoided.
The Krein-Adler transformation (eigenstates deletion based on the
Darboux-Crum transformation) also gives infinitely many exactly
solvable quantum mechanical systems from the original exactly solvable
systems \cite{adler,gos}.
If the original system is described by (ordinary) orthogonal polynomials,
the eigenfunctions of the deformed systems give new orthogonal polynomials.
They also satisfy second order differential or difference equations
and form complete basis. The main difference from the multi-indexed orthogonal
polynomials is their degrees. The set of degrees of these new polynomials is
$\{0,1,2,\ldots\}\backslash\{d_1,d_2,\ldots,d_M\}$, namely there are
several `holes'.
The shape-invariance of the original systems is lost when deformed by the
Krein-Adler transformations, whereas the system retains shape-invariance
when deformed by the method of virtual states deletion.
Since the new orthogonal polynomials obtained by the Krein-Adler
transformation are not so natural in these senses, we are mainly interested
in the multi-indexed orthogonal polynomials.

Some properties of the exceptional and multi-indexed orthogonal polynomials
have been studied \cite{gomez,os16,gomez3,os25}, \cite{os17}--\cite{os27},
\cite{os18}--\cite{hst}, \cite{os29}.
However, there remain various properties to be clarified.
In this paper we focus on the recurrence relations.
The ordinary orthogonal polynomials (`0-indexed' orthogonal polynomials) are
completely characterized by the three term recurrence relations \cite{szego}.
The three term recurrence relations are basic properties of the orthogonal
polynomials and it is important to find the corresponding recurrence
relations of the multi-indexed orthogonal polynomials.
We will show that the $M$-indexed orthogonal polynomials of Laguerre, Jacobi,
Wilson and Askey-Wilson types satisfy the $3+2M$ term recurrence relations
\eqref{RRP}.
The coefficients of the $3+2M$ term recurrence relations are determined
in terms of those of the three term recurrence relations of the original
polynomials.
In order to obtain the whole multi-indexed orthogonal polynomials by using
these $3+2M$ recurrence relations, we have to specify the first $M+1$ members
of the polynomials as inputs, which are severely constrained.
It is expected that similar recurrence relations hold for the multi-indexed
Racah and $q$-Racah polynomials \cite{os26}.

This paper is organized as follows.
The essence of the Darboux-Crum approach to quantum mechanical systems and
the multi-indexed orthogonal polynomials is recapitulated in section
\ref{sec:miop}. Section \ref{sec:RR} is the main part of the paper.
After recalling the three term recurrence relations for the Laguerre,
Jacobi, Wilson and Askey-Wilson orthogonal polynomials, we present the
$3+2M$ term recurrence relations of the multi-indexed orthogonal polynomials.
The initial data for the recurrence relations are discussed in
\S\,\ref{sec:ID}.
The final section is for a summary and comments.
The explicit formulas of the multi-indexed orthogonal polynomials of
Laguerre, Jacobi, Wilson and Askey-Wilson types are presented in Appendix.

\section{Quantum Mechanical Systems and Multi-Indexed\\ Orthogonal Polynomials}
\label{sec:miop}

Not only the construction of the multi-indexed orthogonal polynomials but
also the derivation of the recurrence relations are based on the quantum
mechanical formulation.
See, for example, \cite{os24} for the general introduction.
Here we recapitulate the Darboux-Crum approach to quantum mechanical systems
with a continuous dynamical variable $x$ and the multi-indexed orthogonal
polynomials \cite{darb,crum,adler,os15,gos,os25,os27}.

We consider quantum mechanical systems in one dimension ($x_1<x<x_2$),
\begin{align}
  &\mathcal{H}=\mathcal{A}^{\dagger}\mathcal{A},\quad
  \mathcal{H}\phi_n(x)=\mathcal{E}_n\phi_n(x)\ \ (n\in\mathbb{Z}_{\geq 0}),
  \quad 0=\mathcal{E}_0<\mathcal{E}_1<\mathcal{E}_2<\cdots,\\
  &(\phi_n,\phi_m)\eqdef\int_{x_1}^{x_2}\!dx\,\phi_n(x)^*\phi_m(x)
  =h_n\delta_{nm}\ \ (h_n>0).
\end{align}
Any solution of the Schr\"{o}dinger equation
$\mathcal{H}\tilde{\phi}(x)=\tilde{\mathcal{E}}\tilde{\phi}(x)$,
which need not be square integrable, can be used as a seed solution for
a Darboux-Crum transformation.
Let us take $M$ distinct seed solutions $\{\tilde{\phi}_{\text{v}}(x)\}$,
\begin{equation}
  \mathcal{H}\tilde{\phi}_{\text{v}}(x)
  =\tilde{\mathcal{E}}_{\text{v}}\tilde{\phi}_{\text{v}}(x)
  \ \ (\text{v}=d_1,d_2,\ldots,d_M).
\end{equation}
The $s$-step Darboux-Crum transformation with seed solutions
$\tilde{\phi}_{\text{v}}$ ($\text{v}=d_1,d_2,\ldots,d_s$) gives
\begin{align}
  &\mathcal{H}_{d_1\ldots d_s}\eqdef
  \hat{\mathcal{A}}_{d_1\ldots d_s}\hat{\mathcal{A}}_{d_1\ldots d_s}^{\dagger}
  +\tilde{\mathcal{E}}_{d_s},
  \label{DCs1}\\
  &\phi_{d_1\ldots d_s\,n}(x)\eqdef
  \hat{\mathcal{A}}_{d_1\ldots d_s}\phi_{d_1\ldots d_{s-1}\,n}(x),\quad
  \tilde{\phi}_{d_1\ldots d_s\,\text{v}}(x)\eqdef
  \hat{\mathcal{A}}_{d_1\ldots d_s}
  \tilde{\phi}_{d_1\ldots d_{s-1}\,\text{v}}(x),
  \label{DCs2}\\
  &\mathcal{H}_{d_1\ldots d_s}\phi_{d_1\ldots d_s\,n}(x)
  =\mathcal{E}_n\phi_{d_1\ldots d_s\,n}(x),\quad
  \mathcal{H}_{d_1\ldots d_s}\tilde{\phi}_{d_1\ldots d_s\,\text{v}}(x)
  =\tilde{\mathcal{E}}_{\text{v}}\tilde{\phi}_{d_1\ldots d_s\,\text{v}}(x).
  \label{DCs3}
\end{align}
Here the concrete forms of $\hat{\mathcal{A}}_{d_1\ldots d_s}$ are given
in \eqref{Ahs} and \eqref{Ahs2}, and the eigenfunctions
$\phi_{d_1\ldots d_s\,n}(x)$ and the seed solutions
$\tilde{\phi}_{d_1\ldots d_s\,\text{v}}(x)$ are expressed by using the
Wronskians \eqref{phisn} or Casoratians \eqref{phid1..dsn}.
Since these \eqref{DCs1}--\eqref{DCs3} are shown in algebraic way,
\eqref{DCs3} holds for any range of the coupling constants contained in
the system. However, the Hamiltonian $\mathcal{H}_{d_1\ldots d_s}$ may be
singular in general.
By picking up another seed solution $\tilde{\phi}_{d_{s+1}}$,
the Hamiltonian $\mathcal{H}_{d_1\ldots d_s}$ is rewritten as
$\mathcal{H}_{d_1\ldots d_s}=\hat{A}_{d_1\ldots d_{s+1}}^{\dagger}
\hat{A}_{d_1\ldots d_{s+1}}+\tilde{\mathcal{E}}_{d_{s+1}}$.
After $M$ steps, we obtain
($\mathcal{H}^{[M]}\eqdef\mathcal{H}_{d_1\ldots d_M}$,
$\phi^{[M]}_n(x)\eqdef\phi_{d_1\ldots d_M\,n}(x)$),
\begin{equation}
  \mathcal{H}^{[M]}\phi^{[M]}_n(x)=\mathcal{E}_n\phi^{[M]}_n(x).
\end{equation}
If this Hamiltonian $\mathcal{H}^{[M]}$ is non-singular, we have
\begin{equation}
  (\phi^{[M]}_n,\phi^{[M]}_m)
  =\prod_{j=1}^M(\mathcal{E}_n-\tilde{\mathcal{E}}_{d_j})
  \cdot h_n\delta_{nm},
\end{equation}
and the Hamiltonian can be rewritten in the standard form
$\mathcal{H}_{d_1\ldots d_M}=\mathcal{A}_{d_1\ldots d_M}^{\dagger}
\mathcal{A}_{d_1\ldots d_M}$.
Note that the deformed systems are independent of the orders of deletions
($\phi_{d_1\ldots d_s\,n}(x)$ and $\tilde{\phi}_{d_1\ldots d_s\,\text{v}}(x)$
may change sign).

We know several methods for constructing non-singular Hamiltonian
$\mathcal{H}^{[M]}$.
Here we mention two methods.
(\romannumeral1) [Krein-Adler transformation] \cite{adler,gos}.
As seed solutions, the eigenfunctions are chosen,
$\tilde{\phi}_{\text{v}}(x)=\phi_{\text{v}}(x)$,
$\tilde{\mathcal{E}}_{\text{v}}=\mathcal{E}_{\text{v}}$, and the index set
$\{d_1,\ldots,d_M\}$ is required to satisfy the Krein-Adler conditions
$\prod_{j=1}^M(m-d_j)\geq 0$ ($\forall m\in\mathbb{Z}_{\geq 0}$).
In this case the intermediate Hamiltonians $\mathcal{H}_{d_1\ldots d_s}$
may be singular but the final Hamiltonian $\mathcal{H}^{[M]}$ is non-singular.
The eigenfunctions are $\phi^{[M]}_n(x)$ with
$n\in\mathbb{Z}_{\geq 0}\backslash\{d_1,\ldots,d_M\}$.
Compared to the original system $\mathcal{H}$, $M$ states with energy
$\mathcal{E}_{d_j}$ are missing in the deformed system $\mathcal{H}^{[M]}$.

\noindent
(\romannumeral2) [method of virtual states deletion] \cite{os25,os27}.
As seed solutions, the virtual state wavefunctions are taken.
For the definition of the virtual states, see \cite{os25,os27}.
The Hamiltonian $\mathcal{H}^{[M]}$ is non-singular (The parameter range
may be restricted.) and the eigenfunctions are $\phi^{[M]}_n(x)$ with
$n\in\mathbb{Z}_{\geq 0}$.
The deformed system $\mathcal{H}^{[M]}$ is exactly iso-spectral to the
original system $\mathcal{H}$.
In this case the intermediate Hamiltonians $\mathcal{H}_{d_1\ldots d_s}$
are also non-singular and iso-spectral to the original system.

Let us assume that the eigenfunctions of the original systems in
\S\,\ref{sec:QM} and \S\,\ref{sec:idQM} are of polynomial type:
\begin{equation}
  \phi_n(x)=\phi_0(x)P_n\bigl(\eta(x)\bigr)
  \ \ (n\in\mathbb{Z}_{\geq 0}).
  \label{phinform}
\end{equation}
Here $P_n(\eta)$ is a polynomial of degree $n$ in $\eta$ and $\eta=\eta(x)$
is a certain function of $x$, which is called the sinusoidal coordinate
\cite{os7}.
Then the eigenfunctions of the deformed system are also of polynomial type,
\begin{equation}
  \phi^{[M]}_n(x)=\Psi^{[M]}(x)P^{[M]}_n\bigl(\eta(x)\bigr),
  \label{phiMnform}
\end{equation}
where $P^{[M]}_n(\eta)\eqdef P_{d_1\ldots d_M,n}(\eta)$ is a polynomial
in $\eta$.
They are orthogonal to each other,
\begin{equation}
  \int_{x_1}^{x_2}\!dx\,\Psi^{[M]}(x)^2P^{[M]}_n\bigl(\eta(x)\bigr)
  P^{[M]}_m\bigl(\eta(x)\bigr)
  =\prod_{j=1}^M(\mathcal{E}_n-\tilde{\mathcal{E}}_{d_j})
  \cdot h_n\delta_{nm}.
\end{equation}
The degree of $P^{[M]}_n(\eta)$ is different from $n$.
For (\romannumeral1), the degree of $P^{[M]}_n(\eta)$ is generically $\ell+n$,
where $\ell=\sum_{j=1}^Md_j-\frac12M(M+1)$.
The label $n$ does not take all values in $\mathbb{Z}_{\geq 0}$.
It takes $n\in\mathbb{Z}_{\geq 0}\backslash\{d_1,\ldots,d_M\}$,
namely there are $M$ `holes'.
For (\romannumeral2), the degree of $P^{[M]}_n(\eta)$ is generically $\ell+n$,
where a positive integer $\ell$ is determined by ${\{d_1,\ldots,d_M\}}$,
\eqref{ell}.
The label $n$ takes all values in $\mathbb{Z}_{\geq 0}$ and there are no
`holes'.
For the original systems described by the Laguerre, Jacobi, Wilson and
Askey-Wilson polynomials, we call the obtained polynomials the multi-indexed
orthogonal polynomials for Laguerre, Jacobi, Wilson and Askey-Wilson types.
The eigenfunctions $\phi_{d_1\ldots d_s\,n}(x)$ of the intermediate Hamiltonians
$\mathcal{H}_{d_1\ldots d_s}$ also have the following form:
\begin{equation}
  \phi_{d_1\ldots d_s\,n}(x)
  =\Psi_{d_1\ldots d_s}(x)P_{d_1\ldots d_s,n}\bigl(\eta(x)\bigr)
  \ \ (n\in\mathbb{Z}_{\geq 0}).
  \label{phisnform}
\end{equation}

In the rest of the paper we consider the method (\romannumeral2) only.
The quantum systems to be considered have some parameters (coupling
constants), denoted symbolically by $\bm{\lambda}=(\lambda_1,\lambda_2,\ldots)$,
$\phi_n(x)=\phi_n(x;\bm{\lambda})$, $P_n(\eta)=P_n(\eta;\bm{\lambda})$, etc.
As a quantum mechanical system, the parameter range should be chosen
such that the deformed system $\mathcal{H}^{[M]}$ is non-singular.
In this paper, however, we treat the algebraic aspects of the multi-indexed
orthogonal polynomials and various algebraic relations hold independently of
the parameter ranges. So we do not care much about the parameter ranges.
The explicit forms of the multi-indexed orthogonal polynomials of
Laguerre, Jacobi, Wilson and Askey-Wilson types are relegated to Appendix,
since they are somewhat lengthy and they are not directly needed for the
derivation of the recurrence relations.

\subsection{Ordinary quantum mechanics}
\label{sec:QM}

For ordinary quantum mechanics, the Hamiltonian with zero ground state
energy can be expressed in a factorized form:
\begin{align}
  \mathcal{H}&=-\frac{d^2}{dx^2}+U(x),\quad
  U(x)=\frac{\partial_x^2\phi_0(x)}{\phi_0(x)},\n
  &=\mathcal{A}^{\dagger}\mathcal{A},\qquad
  \mathcal{A}\eqdef\frac{d}{dx}-\partial_x\log\bigl|\phi_0(x)\bigr|,\quad
  \mathcal{A}^{\dagger}=-\frac{d}{dx}-\partial_x\log\bigl|\phi_0(x)\bigr|.
\end{align}
The intertwining operators $\hat{\mathcal{A}}_{d_1\ldots d_s}$ and
$\hat{\mathcal{A}}_{d_1\ldots d_s}^{\dagger}$ are given by
\begin{equation}
  \hat{\mathcal{A}}_{d_1\ldots d_s}\eqdef\frac{d}{dx}
  -\partial_x\log\bigl|\tilde{\phi}_{d_1\ldots d_s}(x)\bigr|,\quad
  \hat{\mathcal{A}}_{d_1\ldots d_s}^{\dagger}=-\frac{d}{dx}
  -\partial_x\log\bigl|\tilde{\phi}_{d_1\ldots d_s}(x)\bigr|,
  \label{Ahs}
\end{equation}
and the eigenfunctions $\phi_{d_1\ldots d_s\,n}(x)$ and the seed solutions
$\tilde{\phi}_{d_1\ldots d_s\,\text{v}}(x)$ have fractional expressions
\begin{equation}
  \phi_{d_1\ldots d_s\,n}(x)
  =\frac{\text{W}[\tilde{\phi}_{d_1},\ldots,\tilde{\phi}_{d_s},\phi_n](x)}
  {\text{W}[\tilde{\phi}_{d_1},\ldots,\tilde{\phi}_{d_s}](x)},\quad
  \tilde{\phi}_{d_1\ldots d_s\,\text{v}}(x)
  =\frac{\text{W}[\tilde{\phi}_{d_1},\ldots,\tilde{\phi}_{d_s},
  \tilde{\phi}_{\text{v}}](x)}
  {\text{W}[\tilde{\phi}_{d_1},\ldots,\tilde{\phi}_{d_s}](x)}.
  \label{phisn}
\end{equation}
Here $\text{W}[f_1,f_2,\ldots,f_n](x)$ is the Wronskian
\begin{equation}
  \text{W}[f_1,f_2,\ldots,f_n](x)\eqdef
  \det\Bigl(\frac{d^{j-1}f_k(x)}{dx^{j-1}}\Bigr)_{1\le j,k\le n},
\end{equation}
and $\text{W}[\cdot](x)=1$ for $n=0$.

We consider two shape-invariant systems, the radial oscillator and
Darboux-P\"{o}schl-Teller potential, whose eigenfunctions are described
by the Laguerre (L) and Jacobi (J) polynomials.
Various data of these systems are:
\begin{align}
  \text{L}:\quad
  &0<x<\infty,\quad
  \bm{\lambda}=g,\quad\bm{\delta}=1,\quad
  g>\frac12,\n
  &U(x;\bm{\lambda})=x^2+\frac{g(g-1)}{x^2}-(1+2g),\quad
  \mathcal{E}_n(\bm{\lambda})=4n,\quad\eta(x)=x^2,\n
  &\phi_0(x;\bm{\lambda})=e^{-\frac12x^2}x^g,\quad
  P_n(\eta;\bm{\lambda})=L_n^{(g-\frac12)}(\eta),\n
  &h_n(\bm{\lambda})=\frac{1}{2\,n!}\,\Gamma(n+g+\tfrac12),\\
  \text{J}:\quad
  &0<x<\frac{\pi}{2},\quad
  \bm{\lambda}=(g,h),\quad\bm{\delta}=(1,1),\quad
  g,h>\frac12,\n
  &U(x;\bm{\lambda})=\frac{g(g-1)}{\sin^2x}
  +\frac{h(h-1)}{\cos^2 x}-(g+h)^2,\quad
  \mathcal{E}_n(\bm{\lambda})=4n(n+g+h),\quad\eta(x)=\cos 2x,\n
  &\phi_0(x;\bm{\lambda})=(\sin x)^g(\cos x)^h,\quad
  P_n(\eta;\bm{\lambda})=P_n^{(g-\frac12,h-\frac12)}(\eta),\n
  &h_n(\bm{\lambda})=\frac{\Gamma(n+g+\frac12)\Gamma(n+h+\frac12)}
  {2\,n!\,(2n+g+h)\Gamma(n+g+h)},
\end{align}
where $L^{(\alpha)}_n(\eta)$ and $P^{(\alpha,\beta)}_n(\eta)$ are the
Laguerre and Jacobi polynomials respectively.

The multi-indexed Laguerre and Jacobi orthogonal polynomials are
given by \eqref{miopLJ}.

\subsection{Discrete quantum mechanics with pure imaginary shifts}
\label{sec:idQM}

The Hamiltonian of the discrete quantum mechanics with pure imaginary
shifts is ($\gamma\in\mathbb{R}$)
\begin{align}
  &\mathcal{H}\eqdef\sqrt{V(x)}\,e^{\gamma p}\sqrt{V^*(x)}
  +\!\sqrt{V^*(x)}\,e^{-\gamma p}\sqrt{V(x)}
  -V(x)-V^*(x)=\mathcal{A}^{\dagger}\mathcal{A},
  \label{H}\\
  &\mathcal{A}\eqdef i\bigl(e^{\frac{\gamma}{2}p}\sqrt{V^*(x)}
  -e^{-\frac{\gamma}{2}p}\sqrt{V(x)}\,\bigr),\quad
  \mathcal{A}^{\dagger}\eqdef -i\bigl(\sqrt{V(x)}\,e^{\frac{\gamma}{2}p}
  -\sqrt{V^*(x)}\,e^{-\frac{\gamma}{2}p}\bigr).
\end{align}
The $*$-operation on an analytic function $f(x)=\sum_na_nx^n$
($a_n\in\mathbb{C}$) is defined by $f^*(x)=\sum_na_n^*x^n$, in which
$a_n^*$ is the complex conjugation of $a_n$.
The eigenfunctions $\phi_n(x)$ and virtual state wavefunctions
$\tilde{\phi}_{\text{v}}(x)$ can be chosen `real', $\phi^*_n(x)=\phi_n(x)$
and $\tilde{\phi}^*_{\text{v}}(x)=\tilde{\phi}_{\text{v}}(x)$.
The intertwining operators $\hat{\mathcal{A}}_{d_1\ldots d_s}$ and
$\hat{\mathcal{A}}_{d_1\ldots d_s}^{\dagger}$ are given by
\begin{align}
  &\hat{\mathcal{A}}_{d_1\ldots d_s}\eqdef
  i\bigl(e^{\frac{\gamma}{2}p}\sqrt{\hat{V}_{d_1\ldots d_s}^*(x)}
  -e^{-\frac{\gamma}{2}p}\sqrt{\hat{V}_{d_1\ldots d_s}(x)}\,\bigr),\n
  &\hat{\mathcal{A}}_{d_1\ldots d_s}^{\dagger}\eqdef
  -i\bigl(\sqrt{\hat{V}_{d_1\ldots d_s}(x)}\,e^{\frac{\gamma}{2}p}
  -\sqrt{\hat{V}_{d_1\ldots d_s}^*(x)}\,e^{-\frac{\gamma}{2}p}\bigr),
  \label{Ahs2}\\
  &\hat{V}_{d_1\ldots d_s}(x)\eqdef
  \sqrt{V(x-i\tfrac{s-1}{2}\gamma)V^*(x-i\tfrac{s+1}{2}\gamma)}\n
  &\phantom{\hat{V}_{d_1\ldots d_s}(x)\eqdef}\times
  \frac{\text{W}_{\gamma}[\tilde{\phi}_{d_1},\ldots,\tilde{\phi}_{d_{s-1}}]
  (x+i\frac{\gamma}{2})}
  {\text{W}_{\gamma}[\tilde{\phi}_{d_1},\ldots,\tilde{\phi}_{d_{s-1}}]
  (x-i\frac{\gamma}{2})}\,
  \frac{\text{W}_{\gamma}[\tilde{\phi}_{d_1},\ldots,\tilde{\phi}_{d_s}]
  (x-i\gamma)}
  {\text{W}_{\gamma}[\tilde{\phi}_{d_1},\ldots,\tilde{\phi}_{d_s}](x)},
  \label{Vhd1..ds}
\end{align}
and the eigenfunctions $\phi_{d_1\ldots d_s\,n}(x)$ and the seed solutions
$\tilde{\phi}_{d_1\ldots d_s\,\text{v}}(x)$ are expressed by
\begin{align}
  &\phi_{d_1\ldots d_s\,n}(x)=A(x)
  \text{W}_{\gamma}[\tilde{\phi}_{d_1},\ldots,\tilde{\phi}_{d_s},\phi_n](x),
  \n
  &\tilde{\phi}_{d_1\ldots d_s\,\text{v}}(x)=A(x)
  \text{W}_{\gamma}[\tilde{\phi}_{d_1},\ldots,\tilde{\phi}_{d_s},
  \tilde{\phi}_{\text{v}}](x),
  \label{phid1..dsn}\\
  &\quad A(x)=\left(
  \frac{\sqrt{\prod_{j=0}^{s-1}V(x+i(\frac{s}{2}-j)\gamma)
  V^*(x-i(\frac{s}{2}-j)\gamma)}}
  {\text{W}_{\gamma}[\tilde{\phi}_{d_1},\ldots,\tilde{\phi}_{d_s}]
  (x-i\frac{\gamma}{2})
  \text{W}_{\gamma}[\tilde{\phi}_{d_1},\ldots,\tilde{\phi}_{d_s}]
  (x+i\frac{\gamma}{2})}\right)^{\frac12}.
  \nonumber
\end{align}
Here $\text{W}_{\gamma}[f_1,f_2,\ldots,f_n](x)$ is the Casoratian
\begin{equation}
  \text{W}_{\gamma}[f_1,\ldots,f_n](x)
  \eqdef i^{\frac12n(n-1)}
  \det\Bigl(f_k\bigl(x^{(n)}_j\bigr)\Bigr)_{1\leq j,k\leq n},\quad
  x_j^{(n)}\eqdef x+i(\tfrac{n+1}{2}-j)\gamma,
  \label{Wdef}
\end{equation}
and $\text{W}_{\gamma}[\cdot](x)=1$ for $n=0$.
We note that the deformed eigenfunctions and seed functions are `real';
$\phi^*_{d_1\ldots d_s\,n}(x)=\phi_{d_1\ldots d_s\,n}(x)$
and $\tilde{\phi}^*_{d_1\ldots d_s\,\text{v}}(x)
=\tilde{\phi}_{d_1\ldots d_s\,\text{v}}(x)$.

We consider two shape-invariant systems whose eigenfunctions are described
by the Wilson (W) and Askey-Wilson (AW) polynomials.
Various data of these systems are:
\begin{align}
  \text{W}:\quad
  &0<x<\infty,\ \ \gamma=1,
  \ \ \bm{\lambda}=(a_1,a_2,a_3,a_4),
  \ \ \bm{\delta}=(\tfrac12,\tfrac12,\tfrac12,\tfrac12),
  \ \ \kappa=1,\n
  &V(x;\bm{\lambda})=\frac{\prod_{j=1}^4(a_j+ix)}{2ix(2ix+1)},\quad
  \mathcal{E}_n(\bm{\lambda})=n(n+b_1-1),
  \ \ b_1\eqdef a_1+a_2+a_3+a_4,\n
  &\phi_0(x;\bm{\lambda})=
  \sqrt{\frac{\prod_{j=1}^4\Gamma(a_j+ix)\Gamma(a_j-ix)}
  {\Gamma(2ix)\Gamma(-2ix)}},
  \qquad\eta(x)=x^2,\ \ \varphi(x)=2x,\n
  &\check{P}_n(x;\bm{\lambda})=P_n\bigl(\eta(x);\bm{\lambda}\bigr)
  =W_n\bigl(\eta(x);a_1,a_2,a_3,a_4\bigr)\n
  &\qquad=(a_1+a_2)_n(a_1+a_3)_n(a_1+a_4)_n
  \ {}_4F_3\Bigl(
  \genfrac{}{}{0pt}{}{-n,\,n+b_1-1,\,a_1+ix,\,a_1-ix}
  {a_1+a_2,\,a_1+a_3,\,a_1+a_4}\Bigm|1\Bigr),\n
  &h_n(\bm{\lambda})=\frac{2\pi n!\,(n+b_1-1)_n
  \prod_{1\leq i<j\leq 4}\Gamma(n+a_i+a_j)}
  {\Gamma(2n+b_1)},\\
  \text{AW}:\quad
  &0<x<\pi,\ \ \gamma=\log q,
  \ \ q^{\bm{\lambda}}=(a_1,a_2,a_3,a_4),
  \ \ \bm{\delta}=(\tfrac12,\tfrac12,\tfrac12,\tfrac12),
  \ \ \kappa=q^{-1},\n
  &V(x;\bm{\lambda})=\frac{\prod_{j=1}^4(1-a_je^{ix})}
  {(1-e^{2ix})(1-qe^{2ix})},\quad
  \mathcal{E}_n(\bm{\lambda})=(q^{-n}-1)(1-b_4q^{n-1}),
  \ \ b_4\eqdef a_1a_2a_3a_4,\n
  &\phi_0(x;\bm{\lambda})=
  \sqrt{\frac{(e^{2ix},e^{-2ix};q)_{\infty}}
  {\prod_{j=1}^4(a_je^{ix},a_je^{-ix};q)_{\infty}}},
  \qquad\eta(x)=\cos x,\ \ \varphi(x)=2\sin x,\n
  &\check{P}_n(x;\bm{\lambda})=P_n\bigl(\eta(x);\bm{\lambda}\bigr)
  =p_n\bigl(\eta(x);a_1,a_2,a_3,a_4|q\bigr)\n
  &\qquad=a_1^{-n}(a_1a_2,a_1a_3,a_1a_4;q)_n
  \ {}_4\phi_3\Bigl(\genfrac{}{}{0pt}{}{q^{-n},\,b_4q^{n-1},\,
  a_1e^{ix},\,a_1e^{-ix}}{a_1a_2,\,a_1a_3,\,a_1a_4}\!\!\Bigm|\!q\,;q\Bigr),\n
  &h_n(\bm{\lambda})=
  \frac{2\pi(b_4q^{n-1};q)_n(b_4q^{2n};q)_{\infty}}
  {(q^{n+1};q)_{\infty}\prod_{1\leq i<j\leq 4}(a_ia_jq^n;q)_{\infty}},
\end{align}
where $W_n(\eta;a_1,a_2,a_3,a_4)$ and $p_n(\eta;a_1,a_2,a_3,a_4|q)$ are
the Wilson and the Askey-Wilson polynomials \cite{koeswart} and
$q^{\bm{\lambda}}$ stands for
$q^{(\lambda_1,\lambda_2,\ldots)}=(q^{\lambda_1},q^{\lambda_2},\ldots)$
and $0<q<1$.
The parameters are restricted by
\begin{equation}
  \{a_1^*,a_2^*,a_3^*,a_4^*\}=\{a_1,a_2,a_3,a_4\}\ \ (\text{as a set});\quad
  \text{W}:\ \text{Re}\,a_i>0,\quad
  \text{AW}:\ |a_i|<1.
  \label{rangeorg}
\end{equation}

The multi-indexed Wilson and Askey-Wilson orthogonal polynomials are
given by \eqref{cPDndef}.

\section{Recurrence Relations}
\label{sec:RR}

Ordinary orthogonal polynomials satisfy the three term recurrence relations
\cite{szego},
\begin{align}
  &\eta P_n(\eta)=A_nP_{n+1}(\eta)+B_nP_n(\eta)+C_nP_{n-1}(\eta),\n
  &\text{or}\quad
  A_nP_{n+1}(\eta)+(B_n-\eta)P_n(\eta)+C_nP_{n-1}(\eta)=0,
  \label{3term}
\end{align}
with $P_{-1}(\eta)\eqdef 0$ and $A_nC_{n+1}>0$ ($n\geq 0$).
When $P_0(\eta)=\text{constant}$ is specified, the entire set of orthogonal
polynomials is determined.
We set
\begin{equation}
  A_{-1}\eqdef0,\quad P_n(\eta)\eqdef0\ \ (n<0).
  \label{Am1=0}
\end{equation}
Note that \eqref{3term} holds for any integer $n\in\mathbb{Z}$, where
$A_n$ ($n\leq -2$), $B_n$ ($n\leq -1$) and $C_n$ ($n\leq 0$) are
arbitrary numbers, e.g. $0$.
We also set 
\begin{equation}
  P_{d_1\ldots d_M,n}(\eta)\eqdef 0\ \ (n<0),\quad
  \phi_{d_1\ldots d_M\,n}(x)\eqdef 0\ \ (n<0),
\end{equation}
for the multi-indexed orthogonal polynomials.

Corresponding to the three term recurrence relations, the multi-indexed
orthogonal polynomials satisfy certain recurrence relations.
We will show that the $M$-indexed orthogonal polynomials of Laguerre, Jacobi,
Wilson and Askey-Wilson types satisfy $3+2M$ term recurrence relations:
\begin{equation}
  \sum_{k=-M-1}^{M+1}R^{[M]}_{n,k}(\eta)P_{d_1\ldots d_M,n+k}(\eta)=0,
  \label{RRP}
\end{equation}
which holds for $n\in\mathbb{Z}$.
Here the coefficients $R^{[M]}_{n,k}(\eta)$ are polynomials of degree
$M+1-|k|$ in $\eta$ and their explicit forms are given by \eqref{Rdef}
for the Laguerre and Jacobi cases in \S\,\ref{sec:RRQM}, and \eqref{Rcdef}
and \eqref{Rdef2} for the Wilson and Askey-Wilson cases in \S\,\ref{sec:RRidQM}.
These coefficients are expressed in terms of the coefficients of the three
term recurrence relations, $A_n$, $B_n$ and $C_n$, as determined recursively
by \eqref{Rdef} or \eqref{Rcdef}.
In other words, the coefficients $R^{[M]}_{n,k}(\eta)$ are independent of
the deformation data $\{\tilde{\phi}_{d_1}\ldots,\tilde{\phi}_{d_M}\}$
except for $M$.
As will be discussed in \S\ref{sec:ID}, the data
$\{\tilde{\phi}_{d_1}\ldots,\tilde{\phi}_{d_M}\}$ are encoded into the initial
data of the first $M+1$ members of the $M$-indexed orthogonal polynomials.

\subsection{Three term recurrence relations}
\label{sec:3tRR}

Here we give the coefficients of the three term recurrence relations
\eqref{3term} for the standard Laguerre, Jacobi, Wilson and Askey-Wilson
polynomials \cite{koeswart} with the input $P_0(\eta)=1$:
\begin{align}
  \text{L}:\quad
  &A_n=-(n+1),\quad B_n=2n+g+\tfrac12,\quad
  C_n=-(n+g-\tfrac12),
  \label{ABCL}\\
  \text{J}:\quad
  &A_n=\frac{2(n+1)(n+g+h)}{(2n+g+h)(2n+g+h+1)},\quad
  B_n=\frac{(h-g)(g+h-1)}{(2n+g+h-1)(2n+g+h+1)},\n
  &C_n=\frac{2(n+g-\frac12)(n+h-\frac12)}{(2n+g+h-1)(2n+g+h)},\\
  \text{W}:\quad
  &A_n=-\frac{n+b_1-1}{(2n+b_1-1)(2n+b_1)},\quad
  C_n=-\frac{n\prod_{1\leq j<k\leq 4}(n+a_j+a_k-1)}
  {(2n+b_1-2)(2n+b_1-1)},\n
  &B_n=\frac{(n+b_1-1)\prod_{k=2}^4(n+a_1+a_k)}
  {(2n+b_1-1)(2n+b_1)}
  +\frac{n\prod_{2\leq j<k\leq 4}(n+a_j+a_k-1)}
  {(2n+b_1-2)(2n+b_1-1)}
  -a_1^2,\\
  \text{AW}:\quad
  &A_n=\frac{1-b_4q^{n-1}}{2(1-b_4q^{2n-1})(1-b_4q^{2n})},\quad
  C_n=\frac{(1-q^n)\prod_{1\leq j<k\leq 4}(1-a_ja_kq^{n-1})}
  {2(1-b_4q^{2n-2})(1-b_4q^{2n-1})},\n
  &B_n=\frac{a_1+a_1^{-1}}{2}
  -\frac{(1-b_4q^{n-1})\prod_{k=2}^4(1-a_1a_kq^n)}
  {2a_1(1-b_4q^{2n-1})(1-b_4q^{2n})}\n
  &\phantom{B_n=}
  -\frac{a_1(1-q^n)\prod_{2\leq j<k\leq 4}(1-a_ja_kq^{n-1})}
  {2(1-b_4q^{2n-2})(1-b_4q^{2n-1})}.
  \label{ABCAW}
\end{align}

\subsection{Multi-indexed Laguerre and Jacobi polynomials}
\label{sec:RRQM}

In this subsection we derive the recurrence relations for the multi-indexed
Laguerre and Jacobi polynomials.
First we note that the operator
$\hat{\mathcal{A}}=\frac{d}{dx}-\partial_x\hat{w}(x)$ acts on a product of
two functions $f(x)\phi(x)$ as
\begin{equation}
  \hat{\mathcal{A}}\bigl(f(x)\phi(x)\bigr)
  =f(x)\hat{\mathcal{A}}\phi(x)+\partial_xf(x)\,\phi(x).
\end{equation}

Let us define $R^{[s]}_{n,k}(\eta)$ ($n,k\in\mathbb{Z}$,
$s\in\mathbb{Z}_{\geq -1}$) as follows:
\begin{align}
  &R^{[s]}_{n,k}(\eta)=0\ \ (|k|>s+1),\quad R^{[-1]}_{n,0}(\eta)=1,\n
  &R^{[s]}_{n,k}(\eta)=A_nR^{[s-1]}_{n+1,k-1}(\eta)
  +(B_n-\eta)R^{[s-1]}_{n,k}(\eta)+C_nR^{[s-1]}_{n-1,k+1}(\eta)
  \ \ (s\geq 0).
  \label{Rdef}
\end{align}
For example, $R^{[s]}_{n,k}(\eta)$ for $s=0,1$ are
\begin{align*}
  s=0:\quad&R^{[0]}_{n,1}(\eta)=A_n,
  \ \ R^{[0]}_{n,0}(\eta)=B_n-\eta,
  \ \ R^{[0]}_{n,-1}(\eta)=C_n,\\
  s=1:\quad&R^{[1]}_{n,2}(\eta)=A_nA_{n+1},
  \ \  R^{[1]}_{n,1}(\eta)=A_n(B_n+B_{n+1}-2\eta),\\
  &R^{[1]}_{n,0}(\eta)=A_nC_{n+1}+A_{n-1}C_n+(B_n-\eta)^2,\\
  &R^{[1]}_{n,-2}(\eta)=C_nC_{n-1},
  \ \  R^{[1]}_{n,-1}(\eta)=C_n(B_n+B_{n-1}-2\eta).
\end{align*}
This $R^{[s]}_{n,k}(\eta)$ ($|k|\leq s+1$) is a polynomial of degree
$s+1-|k|$ in $\eta$.
By induction in $s$, we can show that
\begin{equation}
  \partial_{\eta}R^{[s]}_{n,k}(\eta)=-(s+1)R^{[s-1]}_{n,k}(\eta)
  \ \ (s\geq 0).
  \label{Rprop}
\end{equation}
We will show the $3+2s$ term recurrence relations of
$\phi^{[s]}_n(x)\eqdef\phi_{d_1\ldots d_s\,n}(x)$,
\begin{align}
  &\sum_{k=-s}^sR^{[s-1]}_{n,k}\bigl(\eta(x)\bigr)
  \phi^{[s]}_{n+k}(x)=s!\bigl(\partial_x\eta(x)\bigr)^s\phi_n(x)
  \ \ (s\geq 0),
  \label{RRphi2}\\
  &\sum_{k=-s-1}^{s+1}R^{[s]}_{n,k}\bigl(\eta(x)\bigr)
  \phi^{[s]}_{n+k}(x)=0\ \ (s\geq 0),
  \label{RRphi}
\end{align}
by induction in $s$ (for $n\in\mathbb{Z}$).
Since $\phi^{[s]}_n(x)$ has the form \eqref{phisnform}, this \eqref{RRphi}
means the recurrence relations of the multi-indexed orthogonal polynomials
\eqref{RRP}.

\noindent
\underline{first step} :
For $s=0$, \eqref{RRphi2} is trivial and \eqref{RRphi} is
\begin{equation*}
  A_n\phi_{n+1}(x)+\bigl(B_n-\eta(x)\bigr)\phi_n(x)+C_n\phi_{n-1}(x)=0,
\end{equation*}
which is the three term recurrence relation itself.
Therefore $s=0$ case holds.

\medskip

\noindent
\underline{second step} :
Assume that \eqref{RRphi2}--\eqref{RRphi} hold till $s$ ($s\ge 0$),
we will show that they also hold for $s+1$.

By applying $\hat{\mathcal{A}}_{d_1\ldots d_{s+1}}$ to \eqref{RRphi}
and using \eqref{Rprop}, we obtain
\begin{align*}
  0&=\sum_{k=-s-1}^{s+1}R^{[s]}_{n,k}\bigl(\eta(x)\bigr)\phi^{[s+1]}_{n+k}(x)
  +\partial_x\eta(x)\sum_{k=-s-1}^{s+1}\partial_{\eta}
  R^{[s]}_{n,k}\bigl(\eta(x)\bigr)\,\phi^{[s]}_{n+k}(x),\\
  &=\sum_{k=-s-1}^{s+1}R^{[s]}_{n,k}\bigl(\eta(x)\bigr)\phi^{[s+1]}_{n+k}(x)
  -(s+1)\partial_x\eta(x)\sum_{k=-s}^{s}R^{[s-1]}_{n,k}\bigl(\eta(x)\bigr)
  \,\phi^{[s]}_{n+k}(x).
\end{align*}
The second term can be expressed as
\begin{align*}
  &\quad\sum_{k=-s}^{s}R^{[s-1]}_{n,k}\bigl(\eta(x)\bigr)
  \,\phi^{[s]}_{n+k}(x)\\
  &=\hat{\mathcal{A}}_{d_1\ldots d_{s}}\sum_{k=-s}^{s}
  R^{[s-1]}_{n,k}\bigl(\eta(x)\bigr)\,\phi^{[s-1]}_{n+k}(x)
  -\partial_x\eta(x)\sum_{k=-s}^{s}\partial_{\eta}
  R^{[s-1]}_{n,k}\bigl(\eta(x)\bigr)\,\phi^{[s-1]}_{n+k}(x)\\
  &=s\partial_x\eta(x)\sum_{k=-s+1}^{s-1}R^{[s-2]}_{n,k}\bigl(\eta(x)\bigr)
  \,\phi^{[s-1]}_{n+k}(x)
  =s!\bigl(\partial_x\eta(x)\bigr)^s\phi_n(x),
\end{align*}
where we have used the induction assumptions and \eqref{Rprop}.
Therefore we have
\begin{equation}
  \sum_{k=-s-1}^{s+1}R^{[s]}_{n,k}\bigl(\eta(x)\bigr)\phi^{[s+1]}_{n+k}(x)
  =(s+1)!\bigl(\partial_x\eta(x)\bigr)^{s+1}\phi_n(x),
  \label{sono1}
\end{equation}
which shows \eqref{RRphi2} with $s\to s+1$.
The three term recurrence relations and this \eqref{sono1} imply
\begin{align*}
  0&=(s+1)!\bigl(\partial_x\eta(x)\bigr)^{s+1}
  \Bigl(A_n\phi_{n+1}(x)+\bigl(B_n-\eta(x)\bigr)\phi_n(x)
  +C_n\phi_{n-1}(x)\Bigr)\\
  &=A_n\sum_{k=-s-1}^{s+1}R^{[s]}_{n+1,k}\bigl(\eta(x)\bigr)
  \phi^{[s+1]}_{n+1+k}(x)
  +\bigl(B_n-\eta(x)\bigr)\sum_{k=-s-1}^{s+1}R^{[s]}_{n,k}\bigl(\eta(x)\bigr)
  \phi^{[s+1]}_{n+k}(x)\\
  &\quad
  +C_n\sum_{k=-s-1}^{s+1}R^{[s]}_{n-1,k}\bigl(\eta(x)\bigr)
  \phi^{[s+1]}_{n-1+k}(x)\\
  &=\sum_{k=-s-2}^{s+2}\Bigl(A_nR^{[s]}_{n+1,k-1}\bigl(\eta(x)\bigr)
  +\bigl(B_n-\eta(x)\bigr)R^{[s]}_{n,k}\bigl(\eta(x)\bigr)
  +C_nR^{[s]}_{n-1,k+1}\bigl(\eta(x)\bigr)
  \Bigr)\phi^{[s+1]}_{n+k}(x)\\
  &=\sum_{k=-s-2}^{s+2}R^{[s+1]}_{n,k}\bigl(\eta(x)\bigr)
  \phi^{[s+1]}_{n+k}(x),
\end{align*}
which shows \eqref{RRphi} with $s\to s+1$.
This concludes the induction proof of \eqref{RRphi2}--\eqref{RRphi}.

\subsection{Multi-indexed Wilson and Askey-Wilson polynomials}
\label{sec:RRidQM}

In this subsection we derive the recurrence relations for the multi-indexed
Wilson and Askey-Wilson polynomials.
First we note that the operator
$\hat{\mathcal{A}}=i\bigl(e^{\frac{\gamma}{2}p}\sqrt{\hat{V}^*(x)}
-e^{-\frac{\gamma}{2}p}\sqrt{\hat{V}(x)}\,\bigr)$
acts on a product of two functions $f(x)\phi(x)$ as
\begin{align}
  &\quad\hat{\mathcal{A}}\bigl(f(x)\phi(x)\bigr)\n
  &=i\Bigl(\sqrt{\hat{V}^*(x-i\tfrac{\gamma}{2})}\,f(x-i\tfrac{\gamma}{2})
  \phi(x-i\tfrac{\gamma}{2})
  -\sqrt{\hat{V}(x+i\tfrac{\gamma}{2})}\,f(x+i\tfrac{\gamma}{2})
  \phi(x+i\tfrac{\gamma}{2})\Bigr)\n
  &=f^{(+)}(x)i\Bigl(
  \sqrt{\hat{V}^*(x-i\tfrac{\gamma}{2})}\,\phi(x-i\tfrac{\gamma}{2})
  -\sqrt{\hat{V}(x+i\tfrac{\gamma}{2})}\,\phi(x+i\tfrac{\gamma}{2})\Bigr)\n
  &\quad+f^{(-)}(x)\Bigl(
  \sqrt{\hat{V}^*(x-i\tfrac{\gamma}{2})}\,\phi(x-i\tfrac{\gamma}{2})
  +\sqrt{\hat{V}(x+i\tfrac{\gamma}{2})}\,\phi(x+i\tfrac{\gamma}{2})\Bigr)\n
  &=f^{(+)}(x)\hat{\mathcal{A}}\phi(x)
  +f^{(-)}(x)\Bigl(
  \sqrt{\hat{V}^*(x-i\tfrac{\gamma}{2})}\,\phi(x-i\tfrac{\gamma}{2})
  +\sqrt{\hat{V}(x+i\tfrac{\gamma}{2})}\,\phi(x+i\tfrac{\gamma}{2})\Bigr),
\end{align}
where $f^{(\pm)}(x)$ are defined by
\begin{equation}
  f^{(+)}(x)\eqdef\frac12\bigl(f(x-i\tfrac{\gamma}{2})
  +f(x+i\tfrac{\gamma}{2})\bigr),\quad
  f^{(-)}(x)\eqdef\frac{i}{2}\bigl(f(x-i\tfrac{\gamma}{2})
  -f(x+i\tfrac{\gamma}{2})\bigr).
\end{equation}

Let us define $\check{R}^{[s]}_{n,k}(x)$ ($n,k\in\mathbb{Z}$,
$s\in\mathbb{Z}_{\geq -1}$) as follows:
\begin{align}
  &\check{R}^{[s]}_{n,k}(x)=0\ \ (|k|>s+1),\quad\check{R}^{[-1]}_{n,0}(x)=1,\n
  &\check{R}^{[s]}_{n,k}(x)
  =A_n\check{R}^{[s-1]}_{n+1,k-1}(x+i\tfrac{\gamma}{2})
  +\bigl(B_n-\eta(x-i\tfrac{s}{2}\gamma)\bigr)
  \check{R}^{[s-1]}_{n,k}(x+i\tfrac{\gamma}{2})\n
  &\qquad\qquad
  +C_n\check{R}^{[s-1]}_{n-1,k+1}(x+i\tfrac{\gamma}{2})
  \ \ (s\geq 0).
  \label{Rcdef}
\end{align}
For example, $\check{R}^{[s]}_{n,k}(x)$ for $s=0,1$ are
\begin{align*}
  s=0:\quad&\check{R}^{[0]}_{n,1}(x)=A_n,
  \ \ \check{R}^{[0]}_{n,0}(x)=B_n-\eta(x),
  \ \ \check{R}^{[0]}_{n,-1}(x)=C_n,\\
  s=1:\quad&\check{R}^{[1]}_{n,2}(x)=A_nA_{n+1},
  \ \  \check{R}^{[1]}_{n,1}(x)=A_n\bigl(B_n+B_{n+1}
  -\eta(x-i\tfrac{\gamma}{2})-\eta(x+i\tfrac{\gamma}{2})\bigr),\\
  &\check{R}^{[1]}_{n,0}(x)=A_nC_{n+1}+A_{n-1}C_n
  +\bigl(B_n-\eta(x-i\tfrac{\gamma}{2})\bigr)
  \bigl(B_n-\eta(x+i\tfrac{\gamma}{2})\bigr),\\
  &\check{R}^{[1]}_{n,-2}(x)=C_nC_{n-1},
  \ \  \check{R}^{[1]}_{n,-1}(x)=C_n\bigl(B_n+B_{n-1}
  -\eta(x-i\tfrac{\gamma}{2})-\eta(x+i\tfrac{\gamma}{2})\bigr).
\end{align*}
By induction in $s$, we can show that
\begin{equation}
  \check{R}^{[s]\,(-)}_{n,k}(x)
  =-\frac{i}{2}\bigl(\eta(x-i\tfrac{s+1}{2}\gamma)
  -\eta(x+i\tfrac{s+1}{2}\gamma)\bigr)\check{R}^{[s-1]}_{n,k}(x)
  \ \ (s\geq 0).
  \label{Rprop2}
\end{equation}
By using this, we obtain the following expression of
$\check{R}^{[s]}_{n,k}(x)$:
\begin{align}
  \check{R}^{[s]}_{n,k}(x)&=A_n\check{R}^{[s-1]\,(+)}_{n+1,k-1}(x)
  +\Bigl(B_n-\frac12\bigl(\eta(x-i\tfrac{s}{2}\gamma)
  +\eta(x+i\tfrac{s}{2}\gamma)\bigr)\Bigr)\check{R}^{[s-1]\,(+)}_{n,k}(x)\n
  &\quad+C_n\check{R}^{[s-1]\,(+)}_{n-1,k+1}(x)
  -\frac14\bigl(\eta(x-i\tfrac{s}{2}\gamma)
  -\eta(x+i\tfrac{s}{2}\gamma)\bigr)^2\check{R}^{[s-2]}_{n,k}(x)
  \ \ (s\geq 0).
  \label{Rprop3}
\end{align}
(For $s=0$, the last term vanishes.)
This recurrence relation with respect to $s$ implies that
$\check{R}^{[s]}_{n,k}(x)$ is a polynomial in
$\eta(x-i\frac{m}{2}\gamma)+\eta(x+i\frac{m}{2}\gamma)$ and
$\bigl(\eta(x-i\frac{m}{2}\gamma)-\eta(x+i\frac{m}{2}\gamma)\bigr)^2$
($m=0,1,\ldots,s$).
On the other hand the sinusoidal coordinate $\eta(x)$ satisfies
\begin{align*}
  \eta(x-i\tfrac{m}{2}\gamma)+\eta(x+i\tfrac{m}{2}\gamma)
  &=\left\{
  \begin{array}{ll}
  2\eta(x)-\frac12m^2&:\text{W}\\[2pt]
  (q^{\frac{m}{2}}+q^{-\frac{m}{2}})\eta(x)&:\text{AW}
  \end{array}\right.,\\
  \eta(x-i\tfrac{m}{2}\gamma)\eta(x+i\tfrac{m}{2}\gamma)
  &=\left\{
  \begin{array}{ll}
  \bigl(\eta(x)+\frac14m^2\bigr)^2&:\text{W}\\[2pt]
  \eta(x)^2+\bigl(\frac12(q^{\frac{m}{2}}-q^{-\frac{m}{2}})\bigr)^2&:\text{AW}
  \end{array}\right.,
\end{align*}
which implies that any symmetric polynomial in $\eta(x-i\frac{m}{2}\gamma)$
and $\eta(x+i\frac{m}{2}\gamma)$ is expressed as a polynomial in $\eta(x)$.
Therefore we obtain
\begin{equation}
  \check{R}^{[s]}_{n,k}(x)=R^{[s]}_{n,k}\bigl(\eta(x)\bigr)
  \ \ (|k|\leq s+1):
  \text{a polynomial of degree $s+1-|k|$ in $\eta(x)$}.
  \label{Rdef2}
\end{equation}
We remark that $\check{R}^{[s]\,*}_{n,k}(x)=\check{R}^{[s]}_{n,k}(x)$.

We will show the $3+2s$ term recurrence relations of
$\phi^{[s]}_n(x)\eqdef\phi_{d_1\ldots d_s\,n}(x)$,
\begin{equation}
  \sum_{k=-s-1}^{s+1}\check{R}^{[s]}_{n,k}(x)\phi^{[s]}_{n+k}(x)=0
  \ \ (s\geq 0),
  \label{RRphi3}
\end{equation}
by induction in $s$ (for $n\in\mathbb{Z}$).
Since $\phi^{[s]}_n(x)$ has the form \eqref{phisnform}, this means the
recurrence relations of the multi-indexed orthogonal polynomials \eqref{RRP}.

\noindent
\underline{first step} :
For $s=0$, \eqref{RRphi3} is
\begin{equation*}
  A_n\phi_{n+1}(x)+\bigl(B_n-\eta(x)\bigr)\phi_n(x)+C_n\phi_{n-1}(x)=0,
\end{equation*}
which is the three term recurrence relation itself.
Therefore $s=0$ case holds.

\medskip

\noindent
\underline{second step} :
Assume that \eqref{RRphi3} holds till $s$ ($s\ge 0$),
we will show that it also holds for $s+1$.
Here we use simplified notation
$\hat{V}^{[s+1]}(x)\eqdef\hat{V}_{d_1\ldots d_{s+1}}(x)$.

By applying $\hat{\mathcal{A}}_{d_1\ldots d_{s+1}}$ to \eqref{RRphi3},
we have
\begin{align*}
  0&=\sum_{k=-s-1}^{s+1}\check{R}^{[s]\,(+)}_{n,k}(x)\phi^{[s+1]}_{n+k}(x)\n
  &\quad+\sum_{k=-s-1}^{s+1}\check{R}^{[s]\,(-)}_{n,k}(x)
  \Bigl(\sqrt{\hat{V}^{[s+1]\,*}(x-i\tfrac{\gamma}{2})}\,
  \phi^{[s]}_{n+k}(x-i\tfrac{\gamma}{2})
  +\sqrt{\hat{V}^{[s+1]}(x+i\tfrac{\gamma}{2})}\,
  \phi^{[s]}_{n+k}(x+i\tfrac{\gamma}{2})\Bigr).
\end{align*}
By using \eqref{Rprop2} this is rewritten as
\begin{equation}
  \sum_{k=-s-1}^{s+1}\check{R}^{[s]\,(+)}_{n,k}(x)\phi^{[s+1]}_{n+k}(x)
  =\frac{i}{2}\bigl(\eta(x-i\tfrac{s+1}{2}\gamma)
  -\eta(x+i\tfrac{s+1}{2}\gamma)\bigr)G^{[s+1]}_n(x),
  \label{sono2}
\end{equation}
where
\begin{equation*}
  G^{[s+1]}_n(x)=\sum_{k=-s}^{s}\check{R}^{[s-1]}_{n,k}(x)
  \Bigl(\sqrt{\hat{V}^{[s+1]\,*}(x-i\tfrac{\gamma}{2})}\,
  \phi^{[s]}_{n+k}(x-i\tfrac{\gamma}{2})
  +\sqrt{\hat{V}^{[s+1]}(x+i\tfrac{\gamma}{2})}\,
  \phi^{[s]}_{n+k}(x+i\tfrac{\gamma}{2})\Bigr).
\end{equation*}
Then we have
\begin{align}
  &\quad A_nG^{[s+1]}_{n+1}(x)+\Bigl(B_n-\frac12\bigl(
  \eta(x-i\tfrac{s+1}{2}\gamma)+\eta(x+i\tfrac{s+1}{2}\gamma)
  \bigr)\Bigr)G^{[s+1]}_n(x)
  +C_nG^{[s+1]}_{n-1}(x)\n
  &=\sqrt{\hat{V}^{[s+1]\,*}(x-i\tfrac{\gamma}{2})}\,\biggl(
  A_n\sum_{k=-s}^s\check{R}^{[s-1]}_{n+1,k}(x)
  \phi^{[s]}_{n+1+k}(x-i\tfrac{\gamma}{2})\n
  &\phantom{=\sqrt{\hat{V}^{[s+1]\,*}(x-i\tfrac{\gamma}{2})}}
  +\Bigl(B_n-\frac12\bigl(\eta(x-i\tfrac{s+1}{2}\gamma)
  +\eta(x+i\tfrac{s+1}{2}\gamma)\bigr)\Bigr)
  \sum_{k=-s}^s\check{R}^{[s-1]}_{n,k}(x)
  \phi^{[s]}_{n+k}(x-i\tfrac{\gamma}{2})\n
  &\phantom{=\sqrt{\hat{V}^{[s+1]\,*}(x-i\tfrac{\gamma}{2})}}
  +C_n\sum_{k=-s}^s\check{R}^{[s-1]}_{n-1,k}(x)
  \phi^{[s]}_{n-1+k}(x-i\tfrac{\gamma}{2})
  \biggr)+(\text{c.c.})\n
  &=\sqrt{\hat{V}^{[s+1]\,*}(x-i\tfrac{\gamma}{2})}\,
  \sum_{k=-s-1}^{s+1}\phi^{[s]}_{n+k}(x-i\tfrac{\gamma}{2})\biggl(
  A_n\check{R}^{[s-1]}_{n+1,k-1}(x)\n
  &\qquad\qquad
  +\Bigl(B_n-\frac12\bigl(\eta(x-i\tfrac{s+1}{2}\gamma)
  +\eta(x+i\tfrac{s+1}{2}\gamma)\bigr)\Bigr)
  \check{R}^{[s-1]}_{n,k}(x)
  +C_n\check{R}^{[s-1]}_{n-1,k+1}(x)
  \biggr)+(\text{c.c.})\n
  &=\sqrt{\hat{V}^{[s+1]\,*}(x-i\tfrac{\gamma}{2})}\,
  \sum_{k=-s-1}^{s+1}\phi^{[s]}_{n+k}(x-i\tfrac{\gamma}{2})\biggl(
  \check{R}^{[s]}_{n,k}(x-i\tfrac{\gamma}{2})\n
  &\qquad\qquad
  +\frac12\bigl(\eta(x-i\tfrac{s+1}{2}\gamma)
  -\eta(x+i\tfrac{s+1}{2}\gamma)\bigr)\Bigr)
  \check{R}^{[s-1]}_{n,k}(x)
  \biggr)+(\text{c.c.})\n
  &=\sqrt{\hat{V}^{[s+1]\,*}(x-i\tfrac{\gamma}{2})}\,
  \sum_{k=-s-1}^{s+1}\check{R}^{[s]}_{n,k}(x-i\tfrac{\gamma}{2})
  \phi^{[s]}_{n+k}(x-i\tfrac{\gamma}{2})\n
  &\quad+\sqrt{\hat{V}^{[s+1]}(x+i\tfrac{\gamma}{2})}\,
  \sum_{k=-s-1}^{s+1}\check{R}^{[s]}_{n,k}(x+i\tfrac{\gamma}{2})
  \phi^{[s]}_{n+k}(x+i\tfrac{\gamma}{2})\n
  &\quad-\frac{i}{2}\bigl(\eta(x-i\tfrac{s+1}{2}\gamma)
  -\eta(x+i\tfrac{s+1}{2}\gamma)\bigr)\n
  &\qquad\times
  \sum_{k=-s}^s\check{R}^{[s-1]}_{n,k}(x)
  i\Bigl(\sqrt{\hat{V}^{[s+1]\,*}(x-i\tfrac{\gamma}{2})}\,
  \phi^{[s]}_{n+k}(x-i\tfrac{\gamma}{2})
  -(\sqrt{\hat{V}^{[s+1]}(x+i\tfrac{\gamma}{2})}\,
  \phi^{[s]}_{n+k}(x+i\tfrac{\gamma}{2})\Bigr)\n
  &=-\frac{i}{2}\bigl(\eta(x-i\tfrac{s+1}{2}\gamma)
  -\eta(x+i\tfrac{s+1}{2}\gamma)\bigr)
  \sum_{k=-s}^s\check{R}^{[s-1]}_{n,k}(x)\phi^{[s+1]}_{n+k}(x),
  \label{sono3}
\end{align}
where we have used induction assumption and (c.c.) represents complex
conjugate. From \eqref{sono2} and \eqref{sono3} we obtain
\begin{align*}
  &\quad\frac14\bigl(\eta(x-i\tfrac{s+1}{2}\gamma)
  -\eta(x+i\tfrac{s+1}{2}\gamma)\bigr)^2
  \sum_{k=-s}^s\check{R}^{[s-1]}_{n,k}(x)\phi^{[s+1]}_{n+k}(x)\\
  &=A_n\sum_{k=-s-1}^{s+1}\check{R}^{[s]\,(+)}_{n+1,k}(x)
  \phi^{[s+1]}_{n+1+k}(x)\\
  &\quad+\Bigl(B_n-\frac{1}{2}\bigl(\eta(x-i\tfrac{s+1}{2}\gamma)
  +\eta(x+i\tfrac{s+1}{2}\gamma)\bigr)\Bigr)
  \sum_{k=-s-1}^{s+1}\check{R}^{[s]\,(+)}_{n,k}(x)\phi^{[s+1]}_{n+k}(x)\\
  &\quad+C_n
  \sum_{k=-s-1}^{s+1}\check{R}^{[s]\,(+)}_{n-1,k}(x)\phi^{[s+1]}_{n-1+k}(x),
\end{align*}
namely,
\begin{align*}
  0&=\sum_{k=-s-2}^{s+2}\phi^{[s+1]}_{n+k}(x)
  \biggl(A_n\check{R}^{[s]\,(+)}_{n+1,k-1}(x)
  +\Bigl(B_n-\frac{1}{2}\bigl(\eta(x-i\tfrac{s+1}{2}\gamma)
  +\eta(x+i\tfrac{s+1}{2}\gamma)\bigr)\Bigr)
  \check{R}^{[s]\,(+)}_{n,k}(x)\\
  &\phantom{=\sum_{k=-s-2}^{s+2}\phi^{[s+1]}_{n+1+k}(x)}
  +C_n\check{R}^{[s]\,(+)}_{n-1,k+1}(x)
  -\frac14\bigl(\eta(x-i\tfrac{s+1}{2}\gamma)
  -\eta(x+i\tfrac{s+1}{2}\gamma)\bigr)^2
  \check{R}^{[s-1]}_{n,k}(x)\biggr)\\
  &=\sum_{k=-s-2}^{s+2}\check{R}^{[s+1]}_{n,k}(x)\phi^{[s+1]}_{n+k}(x),
\end{align*}
where we have used \eqref{Rprop3}.
This shows \eqref{RRphi3} with $s\to s+1$.
This concludes the induction proof of \eqref{RRphi3}.

\subsection{Initial data}
\label{sec:ID}

For ordinary orthogonal polynomials, the three term recurrence relations
with an obvious initial data $P_0(\eta)=\text{constant}$
(and $P_{-1}(\eta)=0$) determine the whole polynomials $\{P_n(\eta)\}$
($n=0,1,\ldots$).
For the multi-indexed orthogonal polynomials $\{P_{d_1\ldots d_M,n}(\eta)\}$
($n=0,1,\ldots$) with the derived $3+2M$ term recurrence relations,
the initial data to determine the whole polynomials are the first $M+1$
members of the polynomials, $P_{d_1\ldots d_M,0}(\eta),
P_{d_1\ldots d_M,1}(\eta),\ldots,P_{d_1\ldots d_M,M}(\eta)$.
They are degree $\ell,\ell+1,\ldots,\ell+M$ polynomials in $\eta$ and
they are severely constrained by the input data of
$\{\tilde{\phi}_{d_1}(x),\ldots,\tilde{\phi}_{d_M}(x)\}$.

The $3+2M$ term recurrence relations \eqref{RRP} hold for any integer
$n\in\mathbb{Z}$, which is classified into three cases
(\romannumeral1) $n\leq-M-2$, (\romannumeral2) $-M-1\leq n\leq-1$
and (\romannumeral3) $n\geq 0$.
For case (\romannumeral1), \eqref{RRP} is trivially satisfied because
of $P_{d_1\ldots d_M,m}(\eta)=0$ ($m\leq 0$).
For case (\romannumeral2), \eqref{RRP} is also trivially satisfied due
to the fact
\begin{equation}
  R^{[M]}_{n,k}(\eta)=0\ \ (-M-1\leq n\leq -1,-n\leq k\leq M+1),
  \label{R=0}
\end{equation}
which is a consequence of our choice $A_{-1}=0$.
For case (\romannumeral3), $P_{d_1\ldots d_M,n+M+1}(\eta)$ is determined
by \eqref{RRP} and it is expressed by lower degree polynomials.
As mentioned above, the polynomials $P_{d_1\ldots d_M,n}(\eta)$
($n\geq M+1$) are determined by the $3+2M$ term recurrence relations
\eqref{RRP} with $M+1$ input data
$P_{d_1\ldots d_M,0}(\eta),P_{d_1\ldots d_M,1}(\eta),\ldots,
P_{d_1\ldots d_M,M}(\eta)$.

Note that the input data $P_{d_1\ldots d_M,n}(\eta)$ ($n=0,1,\ldots,M$)
can also be calculated from the data of the lowest degree polynomial at
each intermediate step, $P_{d_1\ldots d_s,0}(\eta)$ ($s=0,1,\ldots,M$).
Since the $3+2M$ term recurrence relations \eqref{RRP} are equivalent to
\begin{equation}
  \sum_{k=-M-1}^{M+1}R^{[M]}_{n,k}\bigl(\eta(x)\bigr)
  \phi_{d_1\ldots d_M\,n+k}(x)=0,
  \label{RRphi4}
\end{equation}
giving the input data $P_{d_1\ldots d_M,n}(\eta)$ ($n=0,1,\ldots,M$)
are equivalent to giving $\phi_{d_1\ldots d_M\,n}(x)$ ($n=0,1,\ldots,M$).
If $\phi_{d_1\ldots d_s\,0}(x)$ ($s=0,1,\ldots,M$) are given,
$\phi_{d_1\ldots d_M\,n}(x)$ ($n=0,1,\ldots,M$) can be calculated
in the following way.
For $s$ ($s=1,2,\ldots,M$ in tern), applying $\hat{A}_{d_1\ldots d_s}$ to
\eqref{RRphi4} with $(M,n)=(s-1,0)$ gives $\phi_{d_1\ldots d_s\,s}(x)$
in terms of already known functions.
Then applying $\hat{A}_{d_1\ldots d_{s+1}}$ to $\phi_{d_1\ldots d_s\,s}(x)$
gives $\phi_{d_1\ldots d_{s+1}\,s}(x)$, and repeating this, and finally we
obtain $\phi_{d_1\ldots d_M\,s}(x)$.
Giving $\phi_{d_1\ldots d_s\,0}(x)$ is equivalent to giving
$P_{d_1\ldots d_s,0}(\eta)$.
Note that $P_{d_1\ldots d_s,0}(\eta;\bm{\lambda})
\propto\Xi_{d_1\ldots d_s}(\eta;\bm{\lambda}+\bm{\delta})$,
which is a consequence of the shape-invariance \cite{os25,os27}.

\section{Summary and Comments}
\label{summary}

The multi-indexed orthogonal polynomials are a new kind of orthogonal
polynomials satisfying second order differential or difference equations,
whose degrees start from a certain positive integer $\ell$ instead of 0,
so that the constraints of Bochner's theorem are avoided.
In this paper we have presented the recurrence relations of the multi-indexed
orthogonal polynomials of Laguerre, Jacobi, Wilson and Askey-Wilson types.
Corresponding to the three term recurrence relations of the ordinary
orthogonal polynomials, the $M$-indexed orthogonal polynomials satisfy
the $3+2M$ term recurrence relations \eqref{RRP}.
Their coefficients are expressed in terms of those of the original three
term recurrence relations.
They are universal in the following sense;
The derivation is based on
(\romannumeral1) the three term recurrence relations \eqref{3term},
(\romannumeral2) the intertwining relations \eqref{DCs2},
(\romannumeral3) the structure of the intertwining operators \eqref{Ahs},
\eqref{Ahs2} and
(\romannumeral4) the formats of the eigenfunctions \eqref{phisnform}.
The explicit expressions of the coefficients of the three term recurrence
relations \eqref{ABCL}--\eqref{ABCAW} and the explicit definitions of the
multi-indexed orthogonal polynomials given in Appendix are not used.
The multi-indexed orthogonal polynomials of Racah and $q$-Racah types
\cite{os26} are not discussed in this paper but the method is applicable
to them, too. We leave this problem to interested readers.

Although we have considered orthogonal polynomials of degrees
$\{\ell,\ell+1,\ldots\}$, the method presented in this paper can be also
applied to orthogonal polynomials of degrees
$\{0,1,\ldots\}\backslash$ $\{d_1,\ldots,d_M\}$.
In fact, the exceptional Hermite polynomials are extensively studied recently
in \cite{ggm} and recurrence relations of the exceptional Hermite polynomials
labeled by $\{0,1,\ldots\}\backslash\{d_1,\ldots,d_M\}$ are obtained
by using the method presented in this paper.
The coefficient polynomials $R^{[M]}_{n,k}(\eta)$ are explicitly expressed
in terms of Hermite polynomials.
It is an interesting problem to find explicit closed forms of
$R^{[M]}_{n,k}(\eta)$ in terms of the original orthogonal polynomials for
other multi-indexed orthogonal polynomials.

The $3+2M$ term recurrence relations need the initial data consisting of the
first $M+1$ members of the polynomials. Namely, when the first $M+1$ members
of the polynomials are given as inputs, the other members of the polynomials
are determined by the $3+2M$ term recurrence relations.
For ordinary orthogonal polynomials (which start at degree 0), three term
recurrence relations always hold and its converse is also true (Favard's
theorem \cite{szego}); {\em i.e.}
polynomials satisfying the three term recurrence relations become
orthogonal polynomials (with respect to a certain inner product).
It is an interesting challenge to formulate the converse of the $3+2M$ term
recurrence relations.
For example, in order that the polynomials determined by the $3+2M$ term
recurrence relations become orthogonal polynomials or satisfy certain second
order differential or difference equations, what conditions should be imposed
on the first $M+1$ members?

The deformed quantum system labeled by an index set
$\mathcal{D}=\{d_1,\ldots,d_M\}$ may be equivalent to another labeled by
a different index set $\mathcal{D}'=\{d'_1,\ldots,d'_{M'}\}$,
which means that the corresponding two multi-indexed orthogonal polynomials
labeled by $\mathcal{D}$ and $\mathcal{D}'$ are proportional.
This has been mentioned in \cite{os25} and generalized in \cite{os29} for
the Laguerre and Jacobi cases.
The same phenomena happen for the Wilson and Askey-Wilson cases \cite{os27}
(and its generalization).
Therefore, if $M'<M$, the $M$-indexed orthogonal polynomials
$P_{\mathcal{D},n}(\eta)$ also satisfy $3+2M'$ term recurrence relations.

The three term recurrence relations for the ordinary orthogonal
polynomials are closely related to the {\em closure relation} between
the Hamiltonian and the sinusoidal coordinate $\eta(x)$, which leads to the
canonical construction of the creation and annihilation operators $a^{(\pm)}$,
$a^{(+)}\phi_n(x)=A_n\phi_{n+1}(x)$,
$a^{(-)}\phi_n(x)=C_n\phi_{n-1}(x)$ \cite{os7}.
By transforming the original creation/annihilation operators $a^{(\pm)}$
in terms of a series of intertwining operators
$\hat{\mathcal{A}}_{d_1\ldots d_s}$,
$\hat{\mathcal{A}}_{d_1\ldots d_s}^\dagger$,
the creation/annihilation operators of the systems of the multi-indexed
orthogonal polynomials $a^{[M](\pm)}$ are obtained:
\begin{align}
  &a^{[M](\pm)}=
  \hat{\mathcal{A}}_{d_1\ldots d_M}\cdots
  \hat{\mathcal{A}}_{d_1d_2}
  \hat{\mathcal{A}}_{d_1}
  a^{(\pm)}
  \frac{\hat{\mathcal{A}}_{d_1}^{\dagger}}
  {\mathcal{H}-\tilde{\mathcal{E}}_{d_1}}
  \frac{\hat{\mathcal{A}}_{d_1d_2}^{\dagger}}
  {\mathcal{H}_{d_1}-\tilde{\mathcal{E}}_{d_2}}\cdots
  \frac{\hat{\mathcal{A}}_{d_1\ldots d_M}^{\dagger}}
  {\mathcal{H}_{d_1\ldots d_{M-1}}-\tilde{\mathcal{E}}_{d_M}},\n
  &a^{[M](+)}\phi^{[M]}_n(x)=A_n\phi^{[M]}_{n+1}(x),\quad
  a^{[M](-)}\phi^{[M]}_n(x)=C_n\phi^{[M]}_{n-1}(x).
\end{align}
It is interesting to see if the $3+2M$ term recurrence relations presented
in this paper would lead to a generalized closure relation between the
deformed Hamiltonian and the sinusoidal coordinate, and if it would give
the above creation/annihilation operators.

Recurrence relations for the exceptional ($M=1$) Laguerre and Jacobi
polynomials have been discussed in \cite{stz} in the context of
bi-spectrality of orthogonal polynomials \cite{GH2}.
We hope that the recurrence relations obtained in this paper will be used
as a starting point to theoretical developments for various problems
involving bispectrality, generalizations of the Jacobi matrix, spectral
theory, existence of a Riemann-Hilbert problem, etc.

\section*{Acknowledgements}
I thank R.\,Sasaki for useful discussion and careful reading of the manuscript.

\bigskip
\appendix
\section{Definitions of the Multi-Indexed Orthogonal\\ Polynomials
of Laguerre, Jacobi, Wilson and\\ Askey-Wilson Types}
\label{sec:app}

For reader's convenience, we present the explicit definitions of the
multi-indexed orthogonal polynomials of Laguerre and Jacobi types \cite{os25}
and Wilson and Askey-Wilson types \cite{os27}, which are obtained by
the method of virtual states deletion.

There are two types of virtual states, type $\I$ and type $\II$, which are
derived by the discrete symmetries of the original Hamiltonian.
We take the set of virtual states for deletion characterized by the degrees
\begin{equation}
  \mathcal{D}=\{d_1,\ldots,d_M\}=\{d^{\I}_1,\ldots,d^{\I}_{M_{\I}},
  d^{\II}_1,\ldots,d^{\II}_{M_{\II}}\}\quad(M=M_{\I}+M_{\II}),
  \label{D}
\end{equation}
and define
\begin{equation}
  \bm{\lambda}^{[M_{\I},M_{\II}]}
  \eqdef\bm{\lambda}+M_{\I}\tilde{\bm{\delta}}_{\I}
  +M_{\II}\tilde{\bm{\delta}}_{\II}.
\end{equation}
The eigenfunctions $\phi^{[M]}_n(x)=\phi_{d_1\ldots d_M\,n}(x)
=\phi_{\mathcal{D}\,n}(x)$ of the deformed system
$\mathcal{H}^{[M]}=\mathcal{H}_{d_1\ldots d_M}$ $=\mathcal{H}_{\mathcal{D}}$
have the following form:
\begin{equation}
  \phi_{\mathcal{D}\,n}(x)=\Psi_{\mathcal{D}}(x)
  P_{\mathcal{D},n}\bigl(\eta(x)\bigr),
  \label{phiDnform}
\end{equation}
where $P_{\mathcal{D},n}(\eta)=P_{d_1\ldots d_M,n}(\eta)$ is the
multi-indexed orthogonal polynomial and the function
$\Psi_{\mathcal{D}}(x)=\Psi_{d_1\ldots d_M}(x)$ is expressed in terms of
the ground state $\phi_0(x)$ and the denominator polynomial
$\Xi_{\mathcal{D}}(\eta)=\Xi_{d_1\ldots d_M}(\eta)$.
The degrees of the denominator polynomial $\Xi_{\mathcal D}(\eta)$ and the
multi-indexed orthogonal polynomial $P_{\mathcal{D},n}(\eta)$ are
generically $\ell$ and $\ell+n$, respectively, in which $\ell$ is given by
\begin{align}
  \ell&\eqdef\sum_{j=1}^{M_{\I}}d_j^{\I}-\frac12 M_{\I}(M_{\I}-1)
  +\sum_{j=1}^{M_{\II}}d_j^{\II}-\frac12M_{\II}(M_{\II}-1)+M_{\I}M_{\II}\n
  &=\sum_{j=1}^Md_j-\frac12M(M-1)+2M_{\I}M_{\II}.
  \label{ell}
\end{align}

\subsection{Multi-indexed Laguerre and Jacobi polynomials}
\label{sec:A_LJ}

Two types of the virtual states are
\begin{align}
  \text{L1}:&\quad
  \tilde{\phi}^{\I}_{\text{v}}(x;\bm{\lambda})
  \eqdef i^{-g}\phi_{\text{v}}(ix;\bm{\lambda}),
  \ \ \xi^{\I}_{\text{v}}(\eta;\bm{\lambda})\eqdef
  P_{\text{v}}(-\eta;\bm{\lambda}),\n
  &\quad\tilde{\delta}^{\I}\eqdef 1,\quad
  \tilde{\mathcal{E}}^{\I}_{\text{v}}(\bm{\lambda})=-4(g+\text{v}+\tfrac12),\\
  \text{L2}:&\quad
  \tilde{\phi}^{\II}_{\text{v}}(x;\bm{\lambda})
  \eqdef\phi_{\text{v}}\bigl(x;\mathfrak{t}(\bm{\lambda})\bigr),
  \ \ \xi^{\II}_{\text{v}}(\eta;\bm{\lambda})\eqdef
  P_{\text{v}}\bigl(\eta;\mathfrak{t}(\bm{\lambda})\bigr),\n
  &\quad\mathfrak{t}(\bm{\lambda})\eqdef 1-g,
  \ \ \tilde{\delta}^{\II}\eqdef-1,\quad
  \tilde{\mathcal{E}}^{\II}_{\text{v}}(\bm{\lambda})=-4(g-\text{v}-\tfrac12),\\
  \text{J1}:&\quad
  \tilde{\phi}^{\I}_{\text{v}}(x;\bm{\lambda})
  \eqdef\phi_{\text{v}}\bigl(x;\mathfrak{t}^{\I}(\bm{\lambda})\bigr),
  \ \ \xi^{\I}_{\text{v}}(\eta;\bm{\lambda})\eqdef
  P_{\text{v}}\bigl(\eta;\mathfrak{t}^{\I}(\bm{\lambda})\bigr),\n
  &\quad\mathfrak{t}^{\I}(\bm{\lambda})\eqdef(g,1-h),
  \ \ \tilde{\delta}^{\I}\eqdef(1,-1),\quad
  \tilde{\mathcal{E}}^{\I}_{\text{v}}(\bm{\lambda})
  =-4(g+\text{v}+\tfrac12)(h-\text{v}-\tfrac12),\\
  \text{J2}:&\quad
  \tilde{\phi}^{\II}_{\text{v}}(x;\bm{\lambda})
  \eqdef\phi_{\text{v}}\bigl(x;\mathfrak{t}^{\II}(\bm{\lambda})\bigr),
  \ \ \xi^{\II}_{\text{v}}(\eta;\bm{\lambda})\eqdef
  P_{\text{v}}\bigl(\eta;\mathfrak{t}^{\II}(\bm{\lambda})\bigr),\n
  &\quad\mathfrak{t}^{\II}(\bm{\lambda})\eqdef(1-g,h),
  \ \ \tilde{\delta}^{\II}\eqdef(-1,1),\quad
  \tilde{\mathcal{E}}^{\II}_{\text{v}}(\bm{\lambda})
  =-4(g-\text{v}-\tfrac12)(h+\text{v}+\tfrac12).
\end{align}
(We have changed the sign of $\tilde{\bm{\delta}}_{\I,\II}$ from those
in \cite{os25}.)
The function $\Psi_{\mathcal{D}}(x)$ in \eqref{phiDnform} is
\begin{equation}
  \Psi_{\mathcal{D}}(x)=\cF^M\psi_{\mathcal{D}}(x;\bm{\lambda}),\quad
  \psi_{\mathcal{D}}(x;\bm{\lambda})
  \eqdef\frac{\phi_0(x;\bm{\lambda}^{[M_{\I},M_{\II}]})}
  {\Xi_{\mathcal{D}}\bigl(\eta(x);\bm{\lambda}\bigr)},\quad
  \cF\eqdef\left\{
  \begin{array}{ll}
  2&:\text{L}\\
  -4&:\text{J}
  \end{array}\right..
\end{equation}
The denominator polynomial $\Xi_{\mathcal{D}}(\eta)$ and the multi-indexed
orthogonal polynomial $P_{\mathcal{D},n}(\eta)$ are defined by the following
Wronskians:
\begin{align}
  &\Xi_{\mathcal{D}}(\eta;\bm{\lambda})\eqdef
  \text{W}[\mu_1,\ldots,\mu_{M_{\I}},\nu_1,\ldots,\nu_{M_{\II}}](\eta)\n
  &\phantom{\Xi_{\mathcal{D}}(\eta;\bm{\lambda})\eqdef}
  \quad\times\left\{
  \begin{array}{ll}
  e^{-M_{\I}\eta}\,\eta^{(M_{\I}+g-\frac12)M_{\II}}&:\text{L}\\[2pt]
  \bigl(\frac{1-\eta}{2}\bigr)^{(M_{\I}+g-\frac12)M_{\II}}
  \bigl(\frac{1+\eta}{2}\bigr)^{(M_{\II}+h-\frac12)M_{\I}}&:\text{J}
  \end{array}\right.,\\
  &P_{\mathcal{D},n}(\eta;\bm{\lambda})\eqdef
  \text{W}[\mu_1,\ldots,\mu_{M_{\I}},\nu_1,\ldots,\nu_{M_{\II}},P_n](\eta)\n
  &\phantom{P_{\mathcal{D},n}(\eta;\bm{\lambda})\eqdef}
  \quad\times\left\{
  \begin{array}{ll}
  e^{-M_{\I}\eta}\,\eta^{(M_{\I}+g+\frac12)M_{\II}}&:\text{L}\\[2pt]
  \bigl(\frac{1-\eta}{2}\bigr)^{(M_{\I}+g+\frac12)M_{\II}}
  \bigl(\frac{1+\eta}{2}\bigr)^{(M_{\II}+h+\frac12)M_{\I}}&:\text{J}
  \end{array}\right.,
  \label{miopLJ}\\
  &\mu_j=\xi_{d_j^{\I}}^{\I}(\eta;\bm{\lambda})\times\left\{
  \begin{array}{ll}
  e^{\eta}&:\text{L}\\
  \bigl(\frac{1+\eta}{2}\bigr)^{\frac12-h}&:\text{J}
  \end{array}\right.,\quad
  \nu_j=\xi_{d_j^{\II}}^{\II}(\eta;\bm{\lambda})\times\left\{
  \begin{array}{ll}
  \eta^{\frac12-g}&:\text{L}\\
  \bigl(\frac{1-\eta}{2}\bigr)^{\frac12-g}&:\text{J}
  \end{array}\right..
\end{align}

\subsection{Multi-indexed Wilson and Askey-Wilson polynomials}
\label{sec:A_WAW}

Two types of the virtual states are
\begin{align}
  \text{type $\I$}:&\quad
  \tilde{\phi}^{\I}_{\text{v}}(x;\bm{\lambda})
  \eqdef\phi_{\text{v}}\bigl(x;\mathfrak{t}^{\I}(\bm{\lambda})\bigr),
  \ \ \xi^{\I}_{\text{v}}(\eta;\bm{\lambda})\eqdef
  P_{\text{v}}\bigl(\eta;\mathfrak{t}^{\I}(\bm{\lambda})\bigr),
  \ \ \check{\xi}^{\I}_{\text{v}}(x;\bm{\lambda})\eqdef
  \xi^{\I}_{\text{v}}\bigl(\eta(x);\bm{\lambda}\bigr),\n
  &\quad
  \mathfrak{t}^{\I}(\bm{\lambda})
  \eqdef(1-\lambda_1,1-\lambda_2,\lambda_3,\lambda_4),
  \ \ \tilde{\delta}^{\I}\eqdef(-\tfrac12,-\tfrac12,\tfrac12,\tfrac12),\n
  &\quad
  \tilde{\mathcal{E}}^{\I}_{\text{v}}(\bm{\lambda})=\left\{
  \begin{array}{ll}
  -(a_1+a_2-\text{v}-1)(a_3+a_4+\text{v})&:\text{W}\\
  -(1-a_1a_2q^{-\text{v}-1})(1-a_3a_4q^{\text{v}})&:\text{AW}
  \end{array}\right.,\\
  \text{type $\II$}:&\quad
  \tilde{\phi}^{\II}_{\text{v}}(x;\bm{\lambda})
  \eqdef\phi_{\text{v}}\bigl(x;\mathfrak{t}^{\II}(\bm{\lambda})\bigr),
  \ \ \xi^{\II}_{\text{v}}(\eta;\bm{\lambda})\eqdef
  P_{\text{v}}\bigl(\eta;\mathfrak{t}^{\II}(\bm{\lambda})\bigr),
  \ \ \check{\xi}^{\II}_{\text{v}}(x;\bm{\lambda})\eqdef
  \xi^{\II}_{\text{v}}\bigl(\eta(x);\bm{\lambda}\bigr),\n
  &\quad
  \mathfrak{t}^{\II}(\bm{\lambda})
  \eqdef(\lambda_1,\lambda_2,1-\lambda_3,1-\lambda_4),
  \ \ \tilde{\delta}^{\II}\eqdef(\tfrac12,\tfrac12,-\tfrac12,-\tfrac12),\n
  &\quad
  \tilde{\mathcal{E}}^{\II}_{\text{v}}(\bm{\lambda})=\left\{
  \begin{array}{ll}
  -(a_3+a_4-\text{v}-1)(a_1+a_2+\text{v})&:\text{W}\\
  -(1-a_3a_4q^{-\text{v}-1})(1-a_1a_2q^{\text{v}})&:\text{AW}
  \end{array}\right..
\end{align}
The function $\Psi_{\mathcal{D}}(x)$ in \eqref{phiDnform} is
\begin{align}
  &\Psi_{\mathcal{D}}(x)=
  \alpha^{\I}(\bm{\lambda}^{[M_{\I},M_{\II}]})^{\frac12M_{\I}}
  \alpha^{\II}(\bm{\lambda}^{[M_{\I},M_{\II}]})^{\frac12M_{\II}}
  \kappa^{-\frac14M_{\I}(M_{\I}+1)-\frac14M_{\II}(M_{\II}+1)
  +\frac52M_{\I}M_{\II}}
  \psi_{\mathcal{D}}(x;\bm{\lambda}),\n
  &\psi_{\mathcal{D}}(x;\bm{\lambda})\eqdef
  \frac{\phi_0(x;\bm{\lambda}^{[M_{\I},M_{\II}]})}
  {\sqrt{\check{\Xi}_{\mathcal{D}}(x-i\frac{\gamma}{2};\bm{\lambda})
  \check{\Xi}_{\mathcal{D}}(x+i\frac{\gamma}{2};\bm{\lambda})}},
\end{align}
where $\alpha^{\I}(\bm{\lambda})$ and $\alpha^{\II}(\bm{\lambda})$ are
\begin{equation}
  \alpha^{\I}(\bm{\lambda})=\left\{
  \begin{array}{ll}
  1&:\text{W}\\
  a_1a_2q^{-1}&:\text{AW}
  \end{array}\right.,\quad
  \alpha^{\II}(\bm{\lambda})=\left\{
  \begin{array}{ll}
  1&:\text{W}\\
  a_3a_4q^{-1}&:\text{AW}
  \end{array}\right..
\end{equation}
The denominator polynomial $\Xi_{\mathcal{D}}(\eta)$ and the multi-indexed
orthogonal polynomial $P_{\mathcal{D},n}(\eta)$ are defined by the following
determinants:
\begin{align}
  &\check{\Xi}_{\mathcal{D}}(x;\bm{\lambda})\eqdef
  \Xi_{\mathcal{D}}\bigl(\eta(x);\bm{\lambda}\bigr),\quad
  \check{P}_{\mathcal{D},n}(x;\bm{\lambda})\eqdef
  P_{\mathcal{D},n}\bigl(\eta(x);\bm{\lambda}\bigr),\\
  &\check{\Xi}_{\mathcal{D}}(x;\bm{\lambda})\eqdef
  A^{-1}\varphi_M(x)^{-1}\,i^{\frac12M(M-1)}\left|
  \begin{array}{llllll}
  \vec{X}^{(M)}_{d^{\I}_1}&\cdots&\vec{X}^{(M)}_{d^{\I}_{M_{\I}}}&
  \vec{Y}^{(M)}_{d^{\II}_1}&\cdots&\vec{Y}^{(M)}_{d^{\II}_{M_{\II}}}\\
  \end{array}\right|,\n
  &\qquad A=\left\{
  \begin{array}{ll}
  \prod_{k=3,4}\prod_{j=1}^{M_{\I}-1}
  (a_k-\frac{M-1}{2}+ix,a_k-\frac{M-1}{2}-ix)_j\\[4pt]
  \ \ \times\prod_{k=1,2}\prod_{j=1}^{M_{\II}-1}
  (a_k-\frac{M-1}{2}+ix,a_k-\frac{M-1}{2}-ix)_j
  &:\text{W}\\[4pt]
  \prod_{k=3,4}\prod_{j=1}^{M_{\I}-1}a_k^{-j}q^{\frac14j(j+1)}
  (a_kq^{-\frac{M-1}{2}}e^{ix},a_kq^{-\frac{M-1}{2}}e^{-ix};q)_j\\[4pt]
  \ \ \times\prod_{k=1,2}\prod_{j=1}^{M_{\II}-1}a_k^{-j}q^{\frac14j(j+1)}
  (a_kq^{-\frac{M-1}{2}}e^{ix},a_kq^{-\frac{M-1}{2}}e^{-ix};q)_j
  &:\text{AW}
  \end{array}\right.,
  \label{cXiDdef}\\[2pt]
  &\check{P}_{\mathcal{D},n}(x;\bm{\lambda})\eqdef
  B^{-1}\varphi_{M+1}(x)^{-1}\n
  &\phantom{\check{P}_{\mathcal{D},n}(x;\bm{\lambda})\eqdef}
  \times i^{\frac12M(M+1)}\left|
  \begin{array}{lllllll}
  \vec{X}^{(M+1)}_{d^{\I}_1}&\cdots&\vec{X}^{(M+1)}_{d^{\I}_{M_{\I}}}&
  \vec{Y}^{(M+1)}_{d^{\II}_1}&\cdots&\vec{Y}^{(M+1)}_{d^{\II}_{M_{\II}}}&
  \vec{Z}^{(M+1)}_n\\
  \end{array}\right|,\n
  &\qquad B=\left\{
  \begin{array}{ll}
  \prod_{k=3,4}\prod_{j=1}^{M_{\I}}
  (a_k-\frac{M}{2}+ix,a_k-\frac{M}{2}-ix)_j\\[4pt]
  \ \ \times\prod_{k=1,2}\prod_{j=1}^{M_{\II}}
  (a_k-\frac{M}{2}+ix,a_k-\frac{M}{2}-ix)_j
  &:\text{W}\\[4pt]
  \prod_{k=3,4}\prod_{j=1}^{M_{\I}}a_k^{-j}q^{\frac14j(j+1)}
  (a_kq^{-\frac{M}{2}}e^{ix},a_kq^{-\frac{M}{2}}e^{-ix};q)_j\\[4pt]
  \ \ \times\prod_{k=1,2}\prod_{j=1}^{M_{\II}}a_k^{-j}q^{\frac14j(j+1)}
  (a_kq^{-\frac{M}{2}}e^{ix},a_kq^{-\frac{M}{2}}e^{-ix};q)_j
  &:\text{AW}
  \end{array}\right.,
  \label{cPDndef}
\end{align}
where
\begin{align}
  &\bigl(\vec{X}^{(M)}_{\text{v}}\bigr)_j
  =r^{\II}_j(x^{(M)}_j;\bm{\lambda},M)
  \check{\xi}^{\I}_{\text{v}}(x^{(M)}_j;\bm{\lambda}),\qquad
  (1\leq j\leq M),\n
  &\bigl(\vec{Y}^{(M)}_{\text{v}}\bigr)_j
  =r^{\I}_j(x^{(M)}_j;\bm{\lambda},M)
  \check{\xi}^{\II}_{\text{v}}(x^{(M)}_j;\bm{\lambda}),\n
  &\bigl(\vec{Z}^{(M)}_n\bigr)_j
  =r^{\II}_j(x^{(M)}_j;\bm{\lambda},M)r^{\I}_j(x^{(M)}_j;\bm{\lambda},M)
  \check{P}_n(x^{(M)}_j;\bm{\lambda}),
\end{align}
and
\begin{align}
  r^{\I}_j(x^{(M)}_j;\bm{\lambda},M)
  &=\alpha^{\I}\bigl(\bm{\lambda}
  +(M-1)\tilde{\bm{\delta}}^{\I}\bigr)^{-\frac12(M-1)}
  \kappa^{\frac12(M-1)^2-(j-1)(M-j)}\\
  &\quad\times\left\{
  \begin{array}{ll}
  {\displaystyle
  \prod_{k=1,2}(a_k-\tfrac{M-1}{2}+ix)_{j-1}(a_k-\tfrac{M-1}{2}-ix)_{M-j}
  }&:\text{W}\\
  {\displaystyle
  e^{ix(M+1-2j)}\prod_{k=1,2}(a_kq^{-\frac{M-1}{2}}e^{ix};q)_{j-1}
  (a_kq^{-\frac{M-1}{2}}e^{-ix};q)_{M-j}
  }&:\text{AW}
  \end{array}\right.,\n
  r^{\II}_j(x^{(M)}_j;\bm{\lambda},M)
  &=\alpha^{\II}\bigl(\bm{\lambda}
  +(M-1)\tilde{\bm{\delta}}^{\II}\bigr)^{-\frac12(M-1)}
  \kappa^{\frac12(M-1)^2-(j-1)(M-j)}\\
  &\quad\times\left\{
  \begin{array}{ll}
  {\displaystyle
  \prod_{k=3,4}(a_k-\tfrac{M-1}{2}+ix)_{j-1}(a_k-\tfrac{M-1}{2}-ix)_{M-j}
  }&:\text{W}\\
  {\displaystyle
  e^{ix(M+1-2j)}\prod_{k=3,4}(a_kq^{-\frac{M-1}{2}}e^{ix};q)_{j-1}
  (a_kq^{-\frac{M-1}{2}}e^{-ix};q)_{M-j}
  }&:\text{AW}
  \end{array}\right..\nonumber
\end{align}
The auxiliary function $\varphi_M(x)$ is defined by
\begin{equation}
  \varphi_M(x)\eqdef
  \varphi(x)^{[\frac{M}{2}]}\prod_{k=1}^{M-2}
  \bigl(\varphi(x-i\tfrac{k}{2}\gamma)\varphi(x+i\tfrac{k}{2}\gamma)
  \bigr)^{[\frac{M-k}{2}]},
\end{equation}
and $\varphi_0(x)=\varphi_1(x)=1$ \cite{gos}.
Here $[x]$ denotes the greatest integer not exceeding $x$.



\begin{thebibliography}{99}
%

\bibitem{infhul}
L.\,Infeld and T.\,E.\,Hull,
``The factorization method,''
Rev. Mod. Phys. {\bf 23} (1951) 21-68.

\bibitem{genden}
L.\,E.\,Gendenshtein,
``Derivation of exact spectra of the Schroedinger equation by means of
supersymmetry,''
JETP Lett. {\bf 38} (1983) 356-359.

\bibitem{susyqm}
F.\,Cooper, A.\,Khare and U.\,Sukhatme,
``Supersymmetry and quantum mechanics,''
Phys. Rep. {\bf 251} (1995) 267-385.

\bibitem{gomez}
D.\,G\'{o}mez-Ullate, N.\,Kamran and R.\,Milson,
``An extension of Bochner's problem: exceptional invariant subspaces,''
J. Approx Theory {\bf 162} (2010) 987-1006,
{\tt arXiv:0805.\hspace{0pt}3376[math-ph]};
``An extended class of orthogonal polynomials defined by a
Sturm-Liouville problem,''
J. Math. Anal. Appl. {\bf 359} (2009) 352-367,
{\tt arXiv:0807.3939[math-\hspace{0pt}ph]}.

\bibitem{quesne}
C.\,Quesne,
``Exceptional orthogonal polynomials, exactly solvable potentials
and supersymmetry,''
J. Phys. {\bf A41} (2008) 392001 (6pp),
{\tt arXiv:0807.4087[quant-ph]}.

\bibitem{os16}
S.\,Odake and R.\,Sasaki,
``Infinitely many shape invariant potentials and new orthogonal polynomials,''
Phys. Lett. {\bf B679} (2009) 414-417,
{\tt arXiv:0906.0142[math-ph]};
%
``Another set of infinitely many exceptional ($X_{\ell}$) Laguerre
polynomials,''
Phys. Lett. {\bf B684} (2010) 173-176,
{\tt arXiv:0911.3442[math-ph]}.

\bibitem{gomez3}  
D.\,G\'{o}mez-Ullate, N.\,Kamran and R.\,Milson,
``Two-step Darboux transformations and exceptional Laguerre polynomials,"
J. Math. Anal. Appl. 387 (2012) 410-418,
{\tt arXiv:\hspace{0pt}1103.5724[math-ph]}.

\bibitem{os25}
S.\,Odake and R.\,Sasaki,
``Exactly solvable quantum mechanics and infinite families of
multi-indexed orthogonal polynomials,"
Phys. Lett. {\bf B702} (2011) 164-170,
{\tt arXiv:\hspace{0pt}1105.0508[math-ph]}.

\bibitem{os18}
S.\,Odake and R.\,Sasaki,
``Infinitely many shape invariant potentials and cubic identities
of the Laguerre and Jacobi polynomials,"
J. Math. Phys. {\bf 51} (2010) 053513 (9pp),
 {\tt arXiv:0911.1585[math-ph]}.

\bibitem{hos}
C.-L.\,Ho, S.\,Odake and R.\,Sasaki,
``Properties of the exceptional ($X_{\ell}$) Laguerre and Jacobi
polynomials,''
SIGMA {\bf 7} (2011) 107 (24pp),
{\tt arXiv:0912.5447[math-ph]}.
 
\bibitem{gkm10}
D.\,G\'{o}mez-Ullate, N.\,Kamran and R.\,Milson,
``Exceptional orthogonal polynomials and the Darboux transformation,"
J. Phys. {\bf A43} (2010) 434016,
{\tt arXiv:1002.2666[math-\hspace{0pt}ph]}.

\bibitem{stz}
R.\,Sasaki, S.\,Tsujimoto and A.\,Zhedanov,
``Exceptional Laguerre and Jacobi polynomials and the corresponding
potentials through Darboux-Crum transformations,"
J. Phys. {\bf A43} (2010) 315204,
{\tt arXiv:1004.4711[math-ph]}.

\bibitem{gomez12}  
D.\,G\'{o}mez-Ullate, N.\,Kamran and R.\,Milson,
``On orthogonal polynomials spanning a non-standard flag,"
Contemp. Math. {\bf 563} (2011) 51-72,
{\tt arXiv:1101.5584[math-ph]}.

\bibitem{gomez4}  
D.\,G\'{o}mez-Ullate, N.\,Kamran and R.\,Milson,
``A conjecture on exceptional orthogonal polynomials,"
Found. Comput. Math. In press,
{\tt arXiv:1203.6857[math-ph]}.

\bibitem{hs4}
C-L.\,Ho and R.\,Sasaki,
``Zeros of the exceptional Laguerre and Jacobi polynomials,"
ISRN Mathematical Physics Volume 2012 (2012) Article ID 920475 (27pp),
{\tt arXiv:\hspace{0pt}1102.5669[math-ph]}.

\bibitem{gmm}
D.\,G\'{o}mez-Ullate, F.\,Marcell\'{a}n and R.\,Milson,
``Asymptotic and interlacing properties of zeros of exceptional Jacobi
and Laguerre polynomials,''
J. Math. Anal. Appl. {\bf 399} (2013) 480-495,
{\tt arXiv:1204.2282[math.CA]}.

\bibitem{st3}
R.\,Sasaki and K.\,Takemura,
``Global solutions of certain second order differential equations with
a high degree of apparent singularity,"
SIGMA {\bf 8} (2012) 085 (18pp),
{\tt arXiv:\hspace{0pt}1207.5302[math.CA]}.

\bibitem{hst}
C.-L.\,Ho, R.\,Sasaki and K.\,Takemura,
``Confluence of apparent singularities in multi-indexed orthogonal
polynomials: the Jacobi case,"
J. Phys. {\bf A46} (2013) 115205 (21pp),
{\tt arXiv:1210.0207[math.CA]}.

\bibitem{bqr}  
B.\,Bagchi, C.\,Quesne and R.\,Roychoudhury,
``Isospectrality of conventional and new extended potentials,
second-order supersymmetry and role of PT symmetry,"
Pramana J. Phys. {\bf 73} (2009) 337-347,
{\tt arXiv:0812.1488[quant-ph]}.

\bibitem{quesne2}  
C.\,Quesne,
``Solvable rational potentials and exceptional orthogonal polynomials
in supersymmetric quantum mechanics,"
SIGMA {\bf 5} (2009) 084 (24pp),
{\tt arXiv:0906.2331\hspace{0pt}[math-ph]}.

\bibitem{duttaroy}  
D.\,Dutta and P.\,Roy,
 ``Conditionally exactly solvable potentials and exceptional orthogonal
polynomials,''
J. Math. Phys. {\bf 51} (2010) 042101 (9pp).

\bibitem{junkroy}  
G.\,Junker and P.\,Roy,
``Conditionally exactly solvable problems and non-linear algebras,"
Phys. Lett. {\bf A 232} (1997) 155-161;
G.\,Junker and P.\,Roy,
``Conditionally Exactly Solvable Potentials:
A Supersymmetric Construction Method,"
Ann. Phys. {\bf 270} (1998) 155-177.

\bibitem{os21}
S.\,Odake and R.\,Sasaki,
``A new family of shape invariantly deformed Darboux-P\"oschl-Teller
potentials with continuous $\ell$,"
J. Phys. {\bf A44} (2011) 195203 (14pp),
{\tt arXiv:1007.\hspace{0pt}3800[math-ph]}.

\bibitem{grandati}  
Y.\,Grandati, 
``Solvable rational extensions of the isotonic oscillator,''
Ann. Phys. {\bf 326} (2011) 2074-2090,
{\tt arXiv:1101.0055[math-ph]};
``Multistep DBT and regular rational extensions of the isotonic oscillator,''
Ann. Phys. {\bf 327} (2012) 2411-2431,
{\tt arXiv:1108.\hspace{0pt}4503[math-ph]}.

\bibitem{grandati2}  
Y.\,Grandati,
``Solvable rational extensions of the Morse and Kepler-Coulomb potentials,"
J. Math. Phys. {\bf 52} (2011) 103505 (12pp),
{\tt arXiv:1103.5023[math-ph]}.

\bibitem{ho1}  
C-L.\,Ho,
``Prepotential approach to solvable rational potentials and exceptional
orthogonal polynomials,"
Prog. Theor. Phys. {\bf 126} (2011) 185-201,
{\tt arXiv:1104.3511[math-ph]}.

\bibitem{ho2}  
C.-L.\,Ho,
``Prepotential approach to solvable rational extensions of harmonic
oscillator and Morse potentials,"
J. Math. Phys. {\bf 52} (2011) 122107 (8pp),
{\tt arXiv:1105.3670\hspace{0pt}[math-ph]}.

\bibitem{quesne4}  
C.\,Quesne,
``Revisiting (quasi-)exactly solvable rational extensions of the Morse
potential,"
Int. J. Mod. Phys. A 27 (2012) 1250073 (18pp),
{\tt arXiv:1203.1812[math-ph]}.

\bibitem{grandati3}  
Y.\,Grandati, 
``New rational extensions of solvable potentials with finite bound state
spectrum,''
Phys. Lett. {\bf A376} (2012) 2866-2872,
{\tt arXiv:1203.4149[math-ph]}.

\bibitem{quesne5}  
C.\,Quesne,
``Novel enlarged shape invariance property and exactly solvable rational
extensions of the Rosen-Morse II and Eckart potentials,"
SIGMA 8 (2012) 080 (19pp),
{\tt arXiv:1208.6165[math-ph]}.

\bibitem{quesne6}  
I.\,Marquette and C.\,Quesne,
``Two-step rational extensions of the harmonic oscillator: exceptional
orthogonal polynomials and ladder operators,"
J. Phys. {\bf A46} (2013) 155201 (14pp),
{\tt arXiv:1212.3474[math-ph]}.

\bibitem{os29}
S.\,Odake and R.\,Sasaki,
``Krein-Adler transformations for shape-invariant potentials and pseudo
virtual states,"
J. Phys. {\bf A46} (2013) 245201 (24pp),
{\tt arXiv:1212.6595[math-\hspace{0pt}ph]}.

\bibitem{os28}
S.\,Odake and R.\,Sasaki,
``Extensions of solvable potentials with finitely many discrete eigenstates,"
J. Phys. {\bf A46} (2013) 235205 (15pp), 
{\tt arXiv:1301.3980[math-ph]}.

\bibitem{os15}
S.\,Odake and R.\,Sasaki,
``Crum's theorem for `discrete' quantum mechanics,''
Prog. Theor. Phys. {\bf 122} (2009) 1067-1079,
{\tt arXiv:0902.2593[math-ph]}.

\bibitem{gos}
L.\,Garc\'ia-Guti\'errez, S.\,Odake and R.\,Sasaki,
``Modification of Crum's theorem for `discrete' quantum mechanics,''
Prog. Theor. Phys. {\bf 124} (2010) 1-26,
{\tt arXiv:1004.0289\hspace{0pt}[math-ph]}.

\bibitem{os24}
S.\,Odake and R.\,Sasaki,
``Discrete quantum mechanics," (Topical Review)
J. Phys. {\bf A44} (2011) 353001 (47pp),
{\tt arXiv:1104.0473[math-ph]}.

\bibitem{os17}
S.\,Odake and R.\,Sasaki,
``Infinitely many shape invariant discrete quantum mechanical systems
and new exceptional orthogonal polynomials related to the Wilson and
Askey-Wilson polynomials,"
Phys. Lett. {\bf B682} (2009) 130-136,
{\tt arXiv:0909.3668[math-ph]}.

\bibitem{os20}
S.\,Odake and R.\,Sasaki,
``Exceptional Askey-Wilson type polynomials through Darboux-Crum
transformations,"
J. Phys. {\bf A43} (2010) 335201 (18pp),
{\tt arXiv:1004.0544[math-\hspace{0pt}ph]}.

\bibitem{os23}
S.\,Odake and R.\,Sasaki,
``Exceptional ($X_{\ell}$) ($q$)-Racah polynomials,"
Prog. Theor. Phys. {\bf 125} (2011) 851-870,
{\tt arXiv:1102.0812[math-ph]}.

\bibitem{os26}
S.\,Odake and R.\,Sasaki,
``Multi-indexed ($q$-)Racah polynomials,"
J. Phys. {\bf A 45} (2012) 385201 (21pp),
{\tt arXiv:1203.5868[math-ph]}.

\bibitem{os27}
S.\,Odake and R.\,Sasaki,
``Multi-indexed Wilson and Askey-Wilson polynomials,"
J. Phys. {\bf A46} (2013) 045204 (22pp),
{\tt arXiv:1207.5584[math-ph]}.

\bibitem{darb}
G.\,Darboux,
{\it Th\'eorie g\'en\'erale des surfaces}
vol 2 (1888) Gauthier-Villars, Paris.

\bibitem{crum}
M.\,M.\,Crum,
``Associated Sturm-Liouville systems,"
Quart. J. Math. Oxford Ser. (2) {\bf 6} (1955) 121-127,
{\tt arXiv:physics/9908019}.

\bibitem{bochner}
E.\,Routh,
``On some properties of certain solutions of a differential equation
of the second order,''
Proc. London Math. Soc. {\bf 16} (1884) 245-261;
S.\,Bochner,
``\"Uber Sturm-Liouvillesche Polynomsysteme,''
Math. Zeit. {\bf 29} (1929) 730-736.

\bibitem{adler}
M.\,G.\,Krein,
``On continuous analogue of a formula of Christoffel from the theory
of orthogonal polynomials," (Russian)
Doklady Acad. Nauk. CCCP, {\bf 113} (1957) 970-973;
V.\,\'E.\,Adler,
``A modification of Crum's method,''
Theor. Math. Phys. {\bf 101} (1994) 1381-1386.

\bibitem{szego}
G.\,Szeg\"o,
{\it Orthogonal polynomials},
Amer. Math. Soc., Providence, RI (1939);
%
T.\,S. Chihara,
{\it An Introduction to Orthogonal Polynomials},
Gordon and Breach, New York (1978);
%
M.\,E.\,H.\,Ismail,
{\it Classical and Quantum Orthogonal Polynomials in One Variable\/},
vol. 98 of Encyclopedia of mathematics and its applications,
Cambridge Univ. Press, Cambridge (2005).

\bibitem{os7}
S.\,Odake and R.\,Sasaki,
``Unified theory of annihilation-creation operators for solvable
(`discrete') quantum mechanics,''
J. Math. Phys. {\bf 47} (2006) 102102 (33pp),
{\tt arXiv:\hspace{0pt}quant-ph/0605215};
``Exact solution in the Heisenberg picture and annihilation-creation
operators,"
Phys. Lett. {\bf B641} (2006) 112-117,
{\tt arXiv:quant-ph/0605221}.

\bibitem{koeswart}
R.\,Koekoek and R.\,F.\,Swarttouw,
``The Askey-scheme of hypergeometric orthogonal polynomials and
its $q$-analogue,''
{\tt arXiv:math.CA/9602214}.

\bibitem{GH2}
F.\,A.\,Gr\"unbaum and L.\,Haine,
``Bispectral Darboux transformations: an extension of the Krall polynomials,''
IMRN (1997) No.\,8, 359--392.

\bibitem{ggm}
D.\,G\'{o}mez-Ullate, Y.\,Grandati and R.\,Milson,
``Rational extensions of the quantum harmonic oscillator and exceptional
Hermite polynomials,''
{\tt arXiv:1306.5143[math-ph]}.

\end{thebibliography}
\end{document}